\documentclass{article}

\pdfoutput=1
\usepackage{arxiv}

\usepackage[utf8]{inputenc} 
\usepackage[T1]{fontenc}    
\usepackage[hidelinks]{hyperref}       
\usepackage{url}            
\usepackage{booktabs}       
\usepackage{amsfonts}       
\usepackage{nicefrac}       
\usepackage{microtype}      
\usepackage{lipsum}
\usepackage{graphicx}
\graphicspath{ {./images/} }
\usepackage{natbib}
\usepackage{amsmath}
\usepackage{makecell}
\usepackage{adjustbox}
\usepackage[table]{xcolor}
\usepackage{enumitem}

\usepackage{multirow}
\usepackage[title]{appendix}
\usepackage[flushleft]{threeparttable}

\def\bSig\mathbf{\Sigma}

\usepackage{natbib}
\usepackage{amsmath}
\usepackage{mathtools}
\usepackage{makecell}
\usepackage{adjustbox}
\usepackage[table]{xcolor}
\usepackage{siunitx}
\usepackage[mathscr]{euscript}
\usepackage{amsfonts}
\usepackage{xcolor}
\usepackage{xr}

\title{Fusing Trial Data for Treatment Comparisons: \\ Single versus Multi-Span Bridging}

\author{Bonnie E. Shook-Sa$^{1*}$, Paul N. Zivich$^{2,3}$, Samuel P. Rosin$^{1}$, Jessie K. Edwards$^{2}$, \\ \textbf{Adaora A. Adimora$^{2,4}$, Michael G. Hudgens$^{1}$, and Stephen R. Cole$^{2}$} \\ 
\\
{$^{1}$Department of Biostatistics, University of North Carolina at Chapel Hill, Chapel Hill, North Carolina, U.S.A.} \\
{$^{2}$Department of Epidemiology, University of North Carolina at Chapel Hill, U.S.A.} \\
{$^{3}$Institute of Global Health and Infectious Diseases, School of Medicine, University of North Carolina at Chapel Hill, U.S.A.} \\
$^{4}$School of Medicine,University of North Carolina at Chapel Hill, U.S.A.  \\ \small{$*$}\footnotesize{bshooksa@email.unc.edu}}

\begin{document}
 \maketitle
 \begin{abstract}
While randomized controlled trials (RCTs) are critical for establishing the efficacy of new therapies, there are limitations regarding what comparisons can be made directly from trial data. RCTs are limited to a small number of comparator arms and often compare a new therapeutic to a standard of care which has already proven efficacious. It is sometimes of interest to estimate the efficacy of the new therapy relative to a treatment that was not evaluated in the same trial, such as a placebo or an alternative therapy that was evaluated in a different trial. Such multi-study comparisons are challenging because of potential differences between trial populations that can affect the outcome. In this paper, two bridging estimators are considered that allow for comparisons of treatments evaluated in different trials using data fusion methods to account for measured differences in trial populations. A ``multi-span'' estimator leverages a shared arm between two trials, while a ``single-span'' estimator does not require a shared arm. A diagnostic statistic that compares the outcome in the standardized shared arms is provided. The two estimators are compared in simulations, where both estimators demonstrate minimal empirical bias and nominal confidence interval coverage when the identification assumptions are met. The estimators are applied to data from the AIDS Clinical Trials Group 320 and 388 to compare the efficacy of two-drug versus four-drug antiretroviral therapy on CD4 cell counts among persons with advanced HIV. The single-span approach requires fewer identification assumptions and was more efficient in simulations and the application.
	\end{abstract}

	
\section{Introduction}\label{sec:Intro}
Randomized controlled trials (RCTs) are essential for establishing the efficacy of new therapies, but there are limitations regarding what comparisons can be made using data from a single trial. For ethical reasons, trials are typically limited to comparisons between a new therapeutic and a standard of care which has already proven efficacious. RCTs often include a small number of comparator arms due to resource constraints. Yet it is sometimes of interest to compare the new therapy to a treatment that was evaluated in a different trial, such as a placebo or an alternative historic therapy \citep{mauri2017challenges}. Such comparisons can be made by making inference across multiple studies. However, simple transitivity arguments can be misleading in this setting, as they ignore differences between trial populations that can affect risk of the outcome \citep{catala2014transitive}. One approach for estimating effects in the setting where two trials share a common arm is network meta-analysis \citep{lumley2002network}. In the setting where the arms of interest are each evaluated in a single (but different) trial, this approach implicitly assumes there are no differences in participant characteristics between trials that affect risk of the outcome. More recently, researchers have estimated placebo counterfactuals, obtaining a standardized estimate of the outcome of interest in the placebo arm of a trial for comparison with an active control in a different trial \citep{Hughes,Donnell}. While this method can account for some differences in trial populations, covariate adjustments are limited to a few discrete covariates. Bayesian methods have also been proposed to estimate placebo counterfactuals but require strong assumptions from expert knowledge or prior data \citep{glidden2020bayesian}. \cite{zhang2016new} present methods for comparing treatments across trials that share a common arm while accounting for differences in trial populations in the point outcome setting using structural nested models and doubly robust methods. \cite{breskin2021fusion} present an inverse probability weighting approach for time-to-event outcomes. 

Bridged treatment comparisons have entailed comparing treatments that were evaluated in separate studies that share common treatment arms \citep{zivich2022bridged}. While previous bridging methods have focused exclusively on the setting where estimators incorporate data from the shared arms (i.e., ``multi-span'' estimators), bridging estimators can also be constructed that ignore data in the shared arms (i.e., ``single-span'' estimators). The Zhang and Breskin estimators, which compare the two arms of interest by anchoring on the common arms, are multi-span bridging estimators. With multi-span estimators, there is an expected difference of zero between standardized shared arms for large samples. This property allows for the construction of a diagnostic test for the estimators. \cite{zivich2022bridged} propose such a diagnostic for the Breskin multi-span estimator in the time-to-event setting. 

Here, a single-span estimator is proposed for bridged treatment comparisons of the intention to treat (ITT) average treatment effect (ATE) that does not leverage the shared arms. This single-span estimator allows inference in the previously described setting but also when there is no shared treatment arm between trials. The single-span estimator uses the same types of adjustments as the multi-span estimator, but only applied to the arms of interest. The single-span estimator requires fewer assumptions than the multi-span estimator because it only requires identification assumptions to hold in trial arms being contrasted and ignores the shared or remaining arms. Multi-span and single-span inverse probability weighting bridging estimators and a bridging diagnostic statistic for the point outcome setting are presented in Section \ref{sec:methods}. The bridging estimators and diagnostic statistic are shown to be consistent and asymptotically normal, and consistent variance estimators are provided. Bias and efficiency of the two bridging approaches are compared by simulations in Section \ref{sec:sims}. In Section \ref{sec:application}, both estimators are applied to compare mean CD4 cell counts after 8 weeks between two- and four-drug antiretroviral (ARV) regimens using data from the AIDS Clinical Trials Group (ACTG) 320 and 388. The benefits and limitations of each approach are discussed in Section \ref{sec:discussion}. The Appendix provides identification proofs for both a single-span and multi-span form of the estimand and proofs of consistency and asymptotic normality for the multi-span and single-span bridging estimators and the bridging diagnostic statistic under the identification assumptions. The Appendix also includes supplemental tables and figures from the simulation study and application. R and Python code is provided on GitHub which implement the proposed methods.

\section{Methods}
\label{sec:methods}
\subsection{Preliminaries}
\label{sec:methods-preliminaries}
Assume that two randomized controlled trials are conducted. In trial one, $n_1$ participants are randomly assigned to treatments $A \in \{1, 2\}$ and in trial two $n_2$ participants are randomized to treatments $A \in \{2, 3\}$, for a total of $n=n_1+n_2$ participants across the two trials. Let $R_i \in \{1,2\}$ indicate that an individual $i$ participated in trial one or two, respectively. Assume trial 2 participants are randomly sampled from the target population, whereas trial 1 participants are sampled from some other (non-focal) population. In both trials, the outcome $Y$ is measured at a single timepoint. Let $Y_i^1$ denote the potential outcome if participant $i$, possibly counter to fact, is assigned treatment $A=1$. Similarly, let $Y_i^2$ and $Y_i^3$ denote potential outcomes for participant $i$ under assigned treatments $A=2$ and $A=3$, respectively, such that under causal consistency $Y_i=\sum_{a=1}^{3} I(A_i=a)Y_i^a$, where in general $I(g)$ is a binary indicator that equals one if $g$ is true and equals zero otherwise. Suppose the outcome $Y$ may be missing (i.e., unobserved) for some individuals. Let the variable $M_i$ indicate whether or not the outcome $Y_i$ was missing, with $M_i=1$ indicating that the outcome was missing and $M_i=0$ otherwise. Assume the vector of covariates $X_{i}$ was collected for all $n$ participants at baseline. Unless noted otherwise, all vectors are assumed to be row vectors. Thus, in trial one, $n_1$ independent and identically distributed (iid) copies of $O_i=\{R_i=1,A_i,M_i,I(M_i=0)Y_i,X_i\}$ are observed with $A_i \in \{1, 2\}$, and in trial two, $n_2$ iid copies of $O_i=\{R_i=2,A_i,M_i,I(M_i=0)Y_i,X_i\}$ are observed with $A_i \in \{2, 3\}$. Observations from participant $i$ in trial one and participant $j$ in trial two are assumed to be independent but not necessarily identically distributed. 

\subsection{Identification Assumptions}
\label{sec:methods-Idassumptions}
Without loss of generality, assume it is of interest to compare assignment to treatments 1 and 3 in the target population, i.e., to estimate ${ATE}^{3-1}=E(Y^3 \mid R=2)-E(Y^1 \mid R=2)$. Note this ITT parameter compares potential outcomes under treatment assignment regardless of adherence to the assigned treatment regimen. The above expression of ${ATE}^{3-1}$ represents the single-span form of the estimand. Alternatively, ${ATE}^{3-1}$ can be rewritten in a multi-span form: $ATE^{3-1}=E(Y^3 \mid R=2)-E(Y^2 \mid R=2)+E(Y^2 \mid R=2)-E(Y^1 \mid R=2)$. Both forms of $ATE^{3-1}$ are identifiable under the assumptions in Table \ref{tab:identcond}. This can be shown by first letting $\pi_A(a,r)=Pr(A=a \mid R=r, X)$, $\pi_M(a,r)=Pr(M=0 \mid A=a, R=r, X)$, and $\pi_R=Pr(R=1 \mid X)$. Note $\pi_A(a,r)$, $\pi_M(a,r)$, and $\pi_R$ depend on $X$ but this is left implicit for notational simplicity. Then under the assumptions in Table 1, for $a \in \{2, 3\}$
\begin{align}
E(Y^a \mid R=2)&=Pr(R=2)^{-1}E\left\{ \frac{1}{\pi_A(a,2)\pi_M(a,2)}I(R=2, A=a, M=0)Y \right \}
 \label{eq:idfintp} \end{align} and likewise, for $a \in \{1, 2\}$, 
	\begin{align}
		E(Y^a \mid R = 2) &= Pr(R=2)^{-1}E\left\{\frac{1-\pi_R}{\pi_R \pi_A(a,1) \pi_M(a,1)} I(R=1, A=a, M=0)Y  \right\}	\label{eq:idfinnfp} \end{align} Proofs of (\ref{eq:idfintp}) and (\ref{eq:idfinnfp}) are provided in the Appendix. Note the expressions on the right of (\ref{eq:idfintp}) and (\ref{eq:idfinnfp}) are identifiable from observed data from the target and non-focal populations, respectively. 
		
\begin{figure}[ht]
\begin{center}
  \includegraphics[width=0.4\columnwidth,trim=1cm 1.25cm 0.5cm 0.1cm,clip]{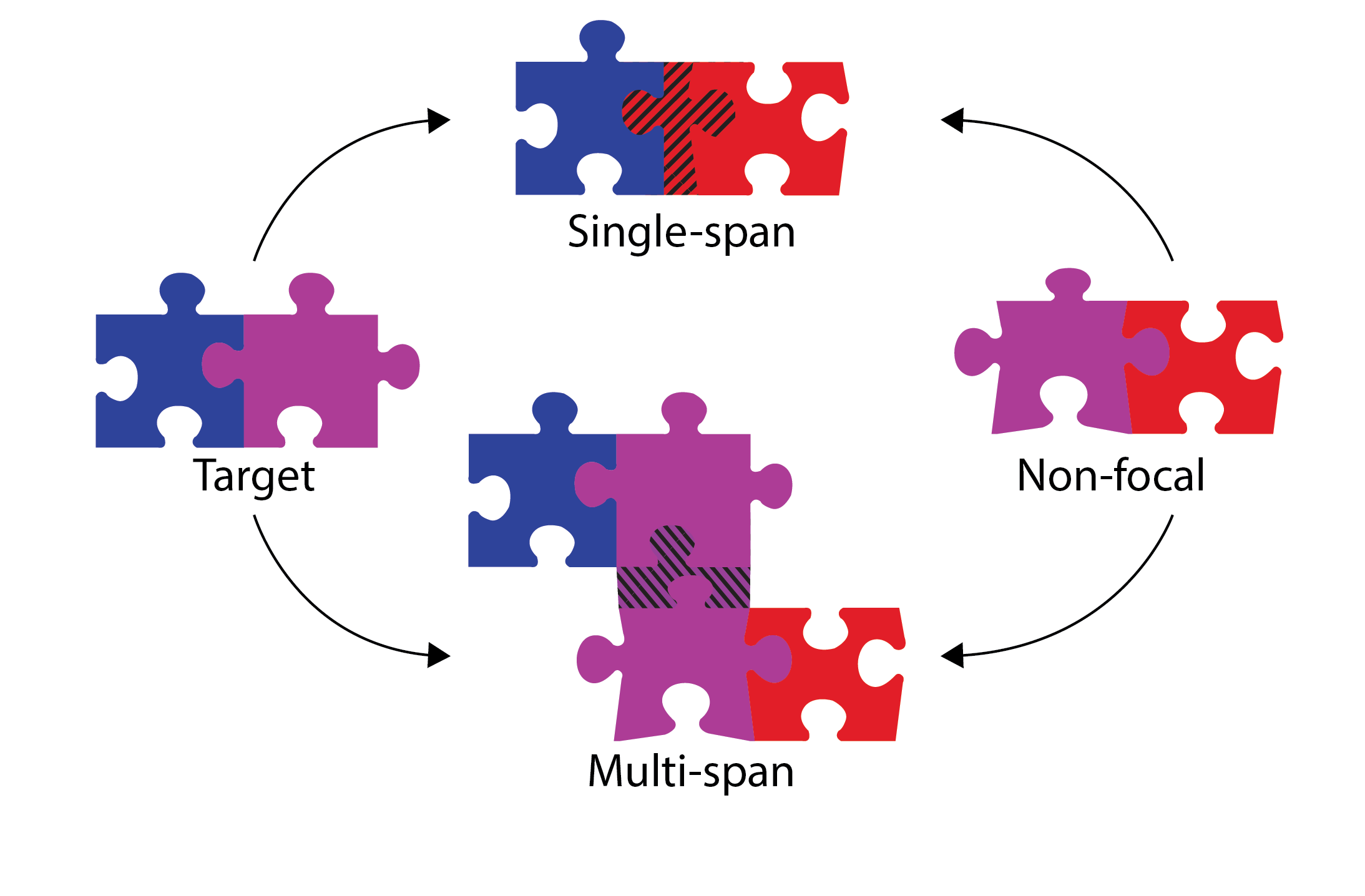}
  \caption{Single-span and multi-span bridged treatment comparisons. The target population is the population to which inference will be made. Here, it is assumed that participants from one trial are a random sample from the target population. The non-focal population is the population from which the other trial was selected.}
  \label{fig:puzzle}
\end{center}
\end{figure}

The single-span and multi-span versions of ${ATE}^{3-1}$ are then identified by contrasting each of their component parts, with $E(Y^3 \mid R=2)$ and $E(Y^1 \mid R=2)$ identified using data from the target and non-focal populations, respectively (Figure \ref{fig:puzzle}). In the multi-span form, $E(Y^2 \mid R=2)$ is identified separately using data from each of the trials, and these terms cancel out such that the multi-span form equals the single-span form. Note from Table \ref{tab:identcond} that the single-span approach requires fewer identification assumptions than the multi-span approach. Namely, it does not require the identification assumptions to hold in the shared arms. This allows for bridging of trials that do not share a common arm or in which (causal) consistency of treatments in the shared arm is questionable, e.g., for trials where different versions of a treatment are used in the shared arm.

\begin{table}
\begin{threeparttable}
\caption{Identification assumptions for the ATE in multi-span and single-span forms}
\label{tab:identcond}
\centering

\begin{tabular}{l l l l} 
\hline
  & & Multi-span & Single-span\\
\hline
\\
1. no measurement error & \makecell[l]{$O$ measured without error \\ for participants with $A=a$} & $a \in \{1, 2, 3\}$ & $a \in \{1, 3\}$ \\ \\
2. causal consistency & $Y=\sum_{a} I(A=a)Y^a$ & $a \in \{1, 2, 3\}$ & $a \in \{1, 3\}$ \\ \\
\makecell[l]{3. conditional exchangeability \\ \hspace{0.4cm}for missingness} & $Y^a \perp M \mid \{A, X, R=r\}$ & \makecell[l]{$a \in \{1, 2\}$ for $r=1$, \\ $a \in \{2, 3\}$ for $r=2$} & \makecell[l]{$a=1$ for $r=1$, \\ $a=3$ for $r=2$} \\ \\
4. positivity for missingness & \makecell[l]{$P(M=0 \mid A=a, X=x, R=r)>0$ \\ for all $a,x$ such that $dF_{AX}(a,x)>0$} & \makecell[l]{$a \in \{1, 2\}$ for $r=1$, \\ $a \in \{2, 3\}$ for $r=2$} & \makecell[l]{$a=1$ for $r=1$, \\ $a=3$ for $r=2$} \\ \\
\makecell[l]{5. conditional exchangeability \\ \hspace{0.4cm}for treatment} & \makecell[l]{$Y^a \perp A \mid \{X, R=r\}$ }& \makecell[l]{$a \in \{1, 2\}$ for $r=1$, \\ $a \in \{2, 3\}$ for $r=2$} & \makecell[l]{$a=1$ for $r=1$, \\ $a=3$ for $r=2$} \\ \\
\makecell[l]{6. positivity for treatment} & \makecell[l]{$P(A=a \mid R=r, X=x)>0$ \\ for all $x,r$ such that $dF_{XR}(x,r)>0$} & \makecell[l]{$a \in \{1, 2\}$ for $r=1$, \\ $a \in \{2, 3\}$ for $r=2$} & \makecell[l]{$a=1$ for $r=1$, \\ $a=3$ for $r=2$} \\ \\
\makecell[l]{7. conditional exchangeability \\ \hspace{0.4cm}for sampling} & \makecell[l]{$Y^a \perp R \mid X$} & $a \in \{1, 2, 3\}$ & $a \in \{1, 3\}$ \\ \\
\makecell[l]{8. positivity for sampling} & \makecell[l]{$P(R=1 \mid X=x)>0$  for all $x$ \\ such that  $dF_{XR}(x, R=2)>0$}   &   &  \\ \\
  \hline 
\end{tabular}
\begin{tablenotes}
      \item Note: $F_{AX}(a,x)$ represents the cumulative distribution function (CDF) of $A,X$ evaluated at $(a,x)$; $F_{XR}(x,r)$ represents the CDF of $X,R$ evaluated at $(x,r)$. 
    \end{tablenotes}
  \end{threeparttable}
\end{table}

\subsection{Estimators}
\label{sec:methods-estimators}
Single and multi-span bridging estimators can be constructed based on the identifiable forms of ${ATE}^{3-1}$ derived in Section \ref{sec:methods-Idassumptions}. For each estimator, the probability of the outcome being non-missing for participant $i$ is estimated using a finite dimensional parametric model. Without loss of generality, consider fitting separate models for each arm within each of the two trials, corresponding to two models for the single-span form (i.e., $A=1, R=1$ and $A=3, R=2$) and four models for the multi-span form (i.e., $A \in \{1, 2\}$ for $R=1$ and $A \in \{2, 3\}$ for $R=2$). For example, the model $\mbox{logit}\{Pr(M_i=0 \mid X_i,A_i=a,R_i=r)\} = g_m(X_i)\lambda_q$ can be fit, where $\lambda_q$ is the column vector of regression coefficients from the missingness model for arm $q$ with $q \in \{1,2,3,4\}$ corresponding to ($A=1, R=1$), ($A=2, R=1$), ($A=2, R=2$), and ($A=3, R=2$), respectively, and $g_m(X_i)$ representing a user-specified function of covariates $X_i$. Inverse-probability of missingness weights are then defined for each participant as $\hat{W}_{Mi}=\{\mbox{expit}(g_m(X_i)\hat{\lambda}_q)\}^{-1}$, where $\hat{\lambda}_q$ is the maximum likelihood estimate (MLE) of $\lambda_q$. Then, inverse-odds of sampling weights \citep{westreich2017transportability} are estimated by regressing the trial indicator $I(R=2)$ on functions of covariates $X$. For example, the model $\mbox{logit}\{Pr(R_i=2 \mid X_i)\} = g_s(X_i)\gamma_e$ can be fit, where $\gamma_e$ is the column vector of regression coefficients from the sampling model with $e \in \{1, 2\}$ corresponding to multi-span and single-span models, respectively, and $g_s(X_i)$ is a user-specified function. The multi-span sampling model incorporates data from $A \in \{1,2,3\}$, while the single-span model only includes data from $A \in \{1,3\}$. Note if the single-span sampling model included data from $A=2$, different identification assumptions would be required than those specified in Table \ref{tab:identcond}. Estimated inverse-odds of sampling weights are then defined for each participant $i$ as $\hat{W}_{Si}=\mbox{exp}\{g_s(X_i)\hat{\gamma}_e\}$, where $\hat{\gamma}_e$ is the MLE for $\gamma_e$. Finally, let ${W}_{Ai}=\{P(A_i=a \mid R_i=r, X_i)\}^{-1}$ represent the reciprocal of the probability that participant $i$ was assigned to treatment $a$ in trial $r \in \{1, 2\}$, conditional on covariates $X_i$. Note $P(A_i=a \mid R_i=r, X_i)$ will usually be known for each trial, but can also be estimated (e.g., from a logistic regression model). Here, it is assumed that $P(A_i=a \mid R_i=r, X_i)$ is known. Assumptions 5 and 6 in Table \ref{tab:identcond} allow randomization to be stratified by $X$, but often randomization is marginal and $Y^a \perp A$ for all $a$ by design. Define the vector of nuisance parameters for the multi-span and single-span estimators as $\eta_1=(\lambda_1,\lambda_2,\lambda_3,\lambda_4,\gamma_1)$ and $\eta_2=(\lambda_1,\lambda_4,\gamma_2)$, respectively.

Multi-span and single-span ITT bridging estimators of $ATE^{3-1}$ can then be constructed as
\begin{equation}\label{est:MS}
    \widehat{ATE}^{3-1}_{m}=\{\hat{E}_2(Y^3 \mid R=2;\hat{\eta}_1) - \hat{E}_2(Y^2 \mid R=2;\hat{\eta}_1)\}+\{\hat{E}_1(Y^2 \mid R=2;\hat{\eta}_1) - \hat{E}_1(Y^1 \mid R=2;\hat{\eta}_1)\}
\end{equation} and 
\begin{equation}\label{est:SS}
    \widehat{ATE}^{3-1}_{s}=\hat{E}_2(Y^3 \mid R=2;\hat{\eta}_2)-\hat{E}_1(Y^1 \mid R=2;\hat{\eta}_2)
\end{equation}
where \begin{equation*}
 \hat{E}_2(Y^a \mid R=2;\hat{\eta}_e)=\frac{\sum_{i=1}^{n}\hat{W}_{Mi}{W}_{Ai}I(A_i=a)I(R_i=2)I(M_i=0)Y_i}{\sum_{i=1}^{n}\hat{W}_{Mi}{W}_{Ai}I(A_i=a)I(R_i=2)I(M_i=0)}
\end{equation*}
and \begin{equation*}
 \hat{E}_1(Y^a \mid R=2;\hat{\eta}_e)=\frac{\sum_{i=1}^{n}\hat{W}_{Mi}\hat{W}_{Si}{W}_{Ai}I(A_i=a)I(R_i=1)I(M_i=0)Y_i}{\sum_{i=1}^{n}\hat{W}_{Mi}\hat{W}_{Si}{W}_{Ai}I(A_i=a)I(R_i=1)I(M_i=0)}
\end{equation*} for $e \in \{1,2\}$ for the multi-span and single-span estimators, respectively. Note $\hat{E}_2(Y^a \mid R=2;\hat{\eta}_e)$ and $\hat{E}_1(Y^a \mid R=2;\hat{\eta}_e)$ are Hajek estimators corresponding to the identifiable forms of the parameters derived in Section \ref{sec:methods-Idassumptions}. The Hajek estimator $\hat{E}_1(Y^a \mid R=2;\hat{\eta}_e)$ is motivated from the identified form of the causal mean (\ref{eq:idfinnfp}) by noting that \begin{align}
		E\left\{\frac{(1-\pi_R)I(A=a)I(R=1)I(M=0)}{\pi_R\pi_A(a,1)\pi_M(a,1)}\right\} &= \Pr(R=2 )	  \label{eq:idfin_denom_haj}
	\end{align} Similarly, the Hajek estimator $\hat{E}_2(Y^a \mid R=2;\hat{\eta}_e)$ is motivated from the identified form of the causal mean (\ref{eq:idfintp}). The proof of (\ref{eq:idfin_denom_haj}) is included in the Appendix.

Both the multi-span (\ref{est:MS}) and single-span (\ref{est:SS}) estimators can be expressed as solutions to vectors of unbiased estimating equations, and are therefore consistent and asymptotically normal estimators of $ATE^{3-1}$ under a set of regularity conditions and the identification assumptions in Section \ref{sec:methods-Idassumptions}. See the Appendix for details. Empirical sandwich variance estimators are obtained using M-estimation \citep{stefanski2002calculus}, stacking the estimating equations from the missingness and sampling models with the estimating equations for the estimators of the counterfactual means $E(Y^a \mid R=2)$ and (\ref{est:MS}) or (\ref{est:SS}). Note if $P(A_i=a \mid R_i=r, X)$ is estimated rather than treated as known then its corresponding estimating equation is stacked as well for the purposes of M-estimation. 

In the multi-span framework, the shared arms are leveraged, providing two estimates of $E(Y^2 \mid R=2)$. Define $D_m=E(Y^2 \mid R=2)-E(Y^2 \mid R=2)=0$ such that in the multi-span form $ATE^{3-1}=E(Y^3 \mid R=2)-E(Y^1 \mid R=2)-D_m$. \cite{zivich2022bridged} propose a diagnostic for bridged treatment comparisons with time-to-event outcomes. A similar diagnostic statistic for underlying assumptions of the multi-span estimator in the point outcome setting is specified as the estimated outcome difference in the shared arms, i.e., \begin{equation}\label{est:diagnostic}
    \widehat{D}_{m}=\hat{E}_2(Y^2 \mid R=2;\hat{\eta}_1)-\hat{E}_1(Y^2 \mid R=2;\hat{\eta}_1)
\end{equation} When the identification assumptions hold, the diagnostic statistic $\widehat{D}_{m}$ can also be expressed as the solution to an unbiased estimating equation such that $\sqrt{n}(\widehat{D}_{m}-D_m)\to^d N(0,V(D))$. The proof of this result is in the Appendix. The asymptotic variance $V(D)$ can be consistently estimated with the empirical sandwich variance estimator and used to construct a Wald-type 95\% confidence interval (CI) for $D_m$. Exclusion of zero from the CI is evidence that at least one of the underlying assumptions of the multi-span estimator is violated. Inclusion of zero in the 95\% CI is supportive of the multi-span estimator, but it does not rule out violation of the assumptions. While this diagnostic evaluates underlying assumptions of the multi-span estimator, in practice the diagnostic may also be used in conjunction with the single-span estimator, as further considered in the discussion. 

\section{Simulation Study}
\label{sec:sims}

Simulation studies were conducted to examine and compare the empirical properties of the bridged estimators and diagnostic statistic described in Section \ref{sec:methods-estimators} along with naive estimators that do not account for differences in trial populations. Naive estimators and a naive diagnostic statistic  $\widetilde{ATE}^{3-1}_{m}$, $\widetilde{ATE}^{3-1}_{s}$, and $\widetilde{D}_{m}$ were equal to (\ref{est:MS}), (\ref{est:SS}), and (\ref{est:diagnostic}), respectively, with $\hat{W}_{Mi}={W}_{Ai}=\hat{W}_{Si}=1$. 

\subsection{Simulation Setup}
\label{sec:sim-setup}
Simulations were designed based on the motivating example in Section \ref{sec:example}, which aimed to compare the effect of two- and four-drug ARV regimens on mean CD4 cell counts at 8 weeks in persons with advanced HIV, controlling for differences between ACTG 320 and 388 trial populations (e.g., differences in distributions of baseline CD4 and history of injection drug use (IDU)). The primary simulations were based on samples of $n_1=1000$ and $n_2=400$ individuals. Simulations were also conducted for $n_1=400$, $n_2=1000$ and for $n_1=1000$, $n_2=2000$, with the results presented in the Appendix. Primary simulations were conducted in R and were independently replicated in Python, with code provided on GitHub. Five scenarios were considered to allow for an examination of the performance of the estimators under different data generating processes. Simulation scenarios included settings where each of the estimators was anticipated to be empirically unbiased as well as settings where they were expected to exhibit bias, as further discussed in Section \ref{sec:sim-results}. 

For each scenario, two covariates $X_1$ and $X_2$ were simulated. Representing an individual's baseline history of IDU, $X_1$ was simulated from a Bernoulli distribution with mean $0.25$ for Scenario 1 and $0.5I(R=1)+0.2I(R=2)$ for Scenarios 2-5. The covariate $X_2$ represented baseline CD4 cell count and was equal to the maximum of 0 and a normally-distributed random variable with mean $\theta$ and standard deviation 30. For Scenario 1, $\theta=175-10X_1$, and for Scenarios 2-5 $\theta=175-20X_1+10I(R=2)$. A Bernoulli random variable with mean 0.5, $B$, was generated. Treatment $A$ was defined as $A=B+1$ for the $n_1$ individuals with $R=1$ such that $A \in \{1, 2\}$, and $A=B+2$ for the $n_2$ individuals with $R=2$ such that $A \in \{2, 3\}$.

Potential outcomes were generated for each individual under the three treatments as follows. First, random variables $U^a$ were generated by taking the maximum of zero and a normally-distributed random variable with mean $\mu^a$ for $a \in \{1, 2, 3\}$ and standard deviation 20, where the values of $\mu^a$ are presented in Table \ref{tab:simscen} for Scenarios 1-5. Then, two types of potential outcomes for treatment $a$ were created: (i) continuous potential outcomes, representing CD4 cell count at 8 weeks, were generated by letting $Y^a=U^a$; (ii) binary potential outcomes, representing whether CD4 cell counts were greater than 250, were generated by letting $Y^a=I(U^a>250)$. For Scenario 1, missing outcome data was induced by generating a Bernoulli random variable $M$ with mean $0.15$. For Scenarios 2-5, the expected value of $M$ was $\mbox{expit}\{0.5X_1-2.0I(R=1)-2.1I(R=2)\}$, corresponding to means of approximately 15\% and 12\% for $R=1$ and $R=2$, respectively. The outcome $Y$ was missing if $M=1$ and observed if $M=0$; outcomes were defined as $Y=Y^1I(A=1)+Y^2I(A=2)+Y^3I(A=3)$ for the continuous and binary cases. 

For each scenario, 2000 simulated samples were generated and $ATE^{3-1}$ was estimated using $\widehat{ATE}^{3-1}_{m}$, $\widehat{ATE}^{3-1}_{s}$, $\widetilde{ATE}^{3-1}_{m}$, and $\widetilde{ATE}^{3-1}_{s}$. In addition, $\widehat{D}_{m}$ and $\widetilde{D}_{m}$ were computed for each sample. Covariate $X_1$ was included in separate logistic regression models fit on each study arm for estimation of inverse probability of missingness weights. Inverse-odds of sampling weights were estimated from logistic regression models with stacked data from the two trials, regressing $R$ on $X_1$ and $X_2$. All observations were included in the sampling model for the multi-span bridging estimator, while only data with $A \in \{1, 3\}$ were included in the single-span model. The asymptotic variance of each estimator was estimated by the empirical sandwich variance estimator using \texttt{geex} in R \citep{saul2020calculus} and \texttt{delicatessen} in Python \citep{zivich2022delicatessen}, and corresponding 95\% Wald CIs were computed. The true $ATE^{3-1}$ was determined empirically by averaging 20 million realizations of $Y^3-Y^1$ from the target population based on the data generating mechanism. The estimand $ATE^{3-1}$ was approximately 60 for the continuous outcome in all scenarios and ranged from 0.45 to 0.51 for the binary outcome across the five scenarios. 

Simulation results were summarized based on empirical bias, average standard error (ASE), empirical standard error (ESE), standard error ratio (ASE/ESE), root mean squared error ($\sqrt{bias^2+ESE^2}$), and empirical 95\% CI coverage, computed as the proportion of simulated samples in which the 95\% CI included the true value of $ATE^{3-1}$. In addition, diagnostic statistics were evaluated by computing the mean value of the diagnostic statistic across the $2000$ simulated samples and the proportion of corresponding 95\% CIs that included zero.

\begin{table}
\begin{threeparttable}
\caption{Conditional mean of continuous potential outcome distributions by simulation scenario}
\label{tab:simscen}
\centering

\begin{tabular}{c c c c} 
\hline
 Scenario & $\mu^1$ & $\mu^2$ & $\mu^3$\\
\hline
\\
1 & $50-5X_1+1.1X_2$ & $80-5X_1+1.1X_2$ & $110-5X_1+1.1X_2$ \\ \\
2 & $35-80X_1+1.0X_2$ & $30-10X_1+1.1X_2$ & $40+20X_1+1.2X_2$ \\ \\
3 & $35-80X_1+1.0X_2$ & \makecell{$I(R=1)\{45-10X_1+1.1X_2\}+$ \\ $I(R=2)\{40+10X_1+1.0X_2\}$} & $40+20X_1+1.2X_2$ \\ \\
4 & $45-80X_1+1.0X_2-30I(R=1)$ & $30-10X_1+1.1X_2$ & $30+20X_1+1.2X_2+20I(R=2)$ \\ \\
5 & $45-80X_1+1.0X_2-30I(R=1)$ & $30-10X_1+1.1X_2+10I(R=1)$ & $30+20X_1+1.2X_2+20I(R=2)$ \\ \\

  \hline 
\end{tabular}
\begin{tablenotes}
      \item Notes: Continuous potential outcomes $Y^a$ were defined as the maximum of zero and a normally-distributed random variable with mean $\mu^a$ for $a \in \{1, 2, 3\}$ and standard deviation 20. Binary potential outcomes were defined as $I(Y^a>250)$ for $a \in \{1, 2, 3\}$. Covariates $X_1$ and $X_2$ are binary and continuous, respectively, and their distributions vary between Scenario 1 and Scenarios 2-5. 
    \end{tablenotes}
  \end{threeparttable}
\end{table}

\subsection{Simulation Results}
\label{sec:sim-results}

\begin{figure}
\begin{center}
  \includegraphics[width=\columnwidth]{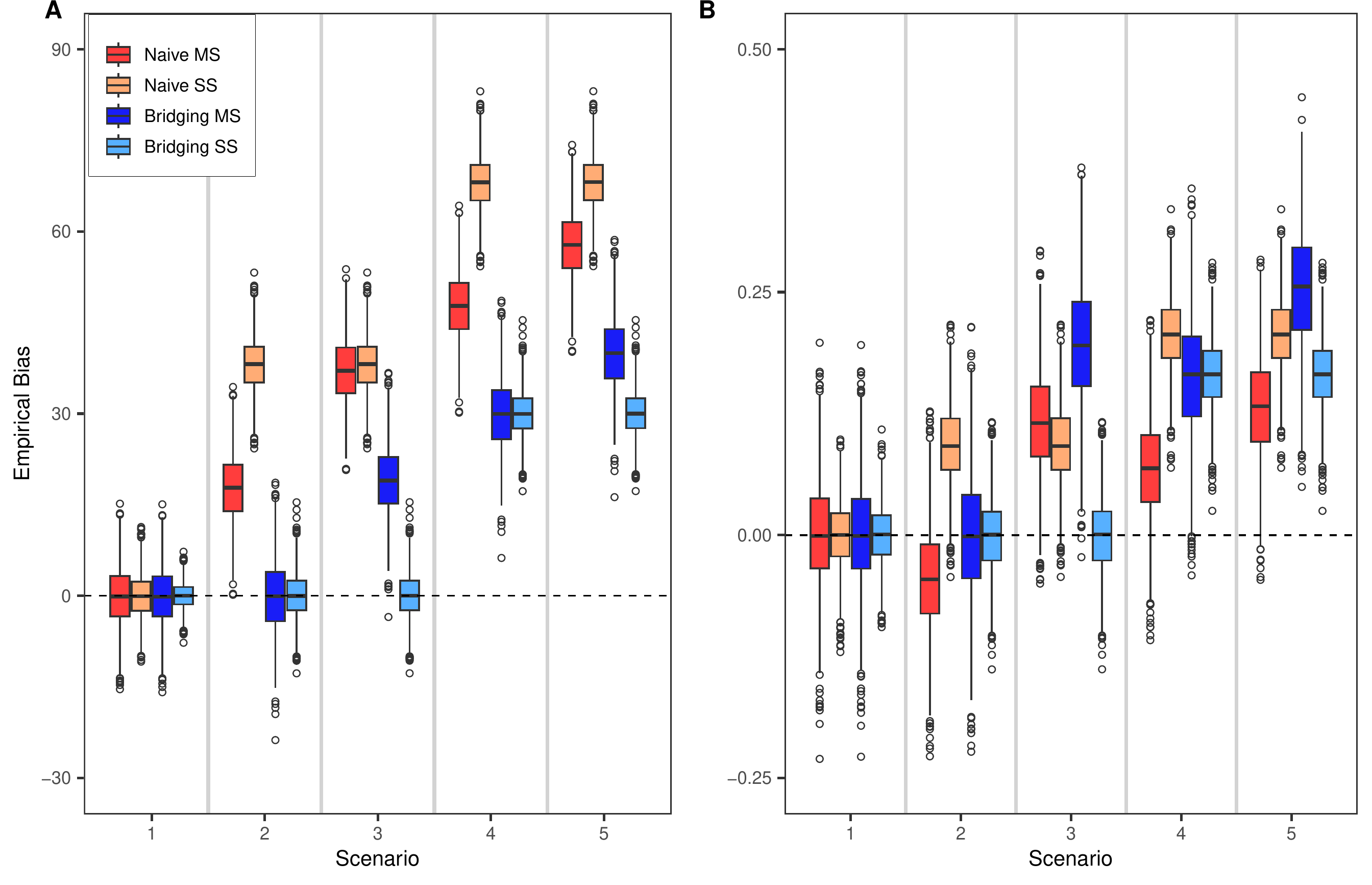}
  \caption{Empirical bias by simulation scenario: (A) continuous outcome (B) binary outcome}
  \label{fig:SimresBias}
\end{center}
\end{figure}

As anticipated, performance of the estimators varied across scenarios. The results of the simulations for $n_1=1000$ and $n_2=400$ are presented in Figure \ref{fig:SimresBias}, with more detailed results in Table \ref{tab:simrescontinuous} and Table \ref{tab:simresbinary} for the continuous and binary outcomes, respectively. In Scenario 1, because the distribution of covariates was the same in the two trial populations and outcome data were missing completely at random, both naive estimators and bridging estimators were empirically unbiased with approximately nominal 95\% CI coverage. In all remaining scenarios, the naive estimators were biased with below nominal CI coverage due to a combination of informative missing data by $X_1$ and effect measure modification by $X_1$ and $X_2$, whose distributions differed between trials. In Scenario 2, informative missingness and effect measure modification by measured covariates was appropriately accounted for by both bridging estimators. Both estimators were empirically unbiased with approximately nominal 95\% CI coverage in this setting. In Scenario 3, a violation of the conditional exchangeability for sampling assumption in the shared arms (i.e., dependence of $\mu^2$ on $R$) led to bias and below nominal CI coverage with the multi-span estimator. Because the single-span estimator does not leverage the shared arms, it was empirically unbiased with nominal CI coverage in this setting. In Scenarios 4 and 5, all estimators exhibited bias and below nominal CI coverage due to a violation of the conditional exchangeability for sampling assumption in the arms of interest (i.e., dependence of $\mu^1$ and $\mu^3$ on $R$). 

Precision of the estimators varied greatly. For all scenarios, the naive single-span and bridging single-span estimators had smaller average and empirical standard errors than their multi-span counterparts. The bridging single-span estimator had the smallest average and empirical standard errors for both outcomes and all scenarios considered. The bridging single-span estimator also had the smallest root mean squared error for all scenarios considered with the continuous outcome and three of the five scenarios for the binary outcome. 

The bridging diagnostic statistic performed as expected in simulations. In Scenarios 3 and 5 where the conditional exchangeability for sampling assumption was violated in the shared arms, the mean diagnostic statistic differed from zero and 100\% and 93\% of diagnostic CIs, respectively, excluded zero for the continuous outcome. For the binary outcome, 100\% and 68\% of diagnostic CIs excluded zero for Scenarios 3 and 5, respectively. In the remaining scenarios where the conditional exchangeability for sampling assumption held in the shared arms, the diagnostic statistic was approximately zero with approximately 95\% of diagnostic CIs including zero. Note in Scenario 4, the conditional exchangeability for sampling assumption held in the shared arms but not in the unshared arms, meaning that the diagnostic failed to detect violation of the multi-span identification assumptions in the arms of interest. This demonstrates how the diagnostic can be used to detect some, but not all, violations of multi-span identification assumptions. Because the naive estimators do not account for differences in trial populations, the naive diagnostic was of limited utility and was potentially misleading in some settings. For example, in Scenario 3, the naive multi-span diagnostic was close to zero with 93\% and 88\% of diagnostic CIs excluding zero for the continuous and binary outcomes, respectively, but there was considerable empirical bias.

The results of the $n_1=400$, $n_2=1000$ and $n_1=1000$, $n_2=2000$ scenarios are presented in Appendix Tables \ref{tab:appn1400cont}-\ref{tab:appn11000bin}. When $n_1=400$ and $n_2=1000$, empirical bias was similar to the setting where $n_1=1000$, $n_2=400$. However, the multi-span and single-span bridging estimators tended to yield larger average estimated standard errors and empirical standard errors, which led to wider CIs and slightly higher 95\% CI coverage in the settings with bias. When $n_1=1000$ and $n_2=2000$, bias was also similar to the $n_1=1000$, $n_2=400$ setting. Due to a larger sample size in the target population trial, average estimated standard errors and empirical standard errors for both bridging estimators were smaller compared to the $n_1=1000$, $n_2=400$ setting, leading to smaller root mean squared errors. Smaller standard errors also resulted in narrower CIs and lower 95\% CI coverage in the scenarios with bias.

In summary, the naive estimators performed poorly in the presence of informative missing outcome data and effect modification based on covariates whose distributions differed between the trials. The bridging estimators performed well when the identification assumptions held, with minimal bias and coverage rates approximately equal to the nominal level for all sample sizes considered. The bridging single-span estimator was much more precise than the multi-span estimator across the scenarios considered and was robust to violations of identification assumptions in the shared arms. The diagnostic statistic performed as anticipated, detecting differences in the outcome between the standardized shared arms.

\begin{table}
\begin{threeparttable}
\caption{Simulation summary results for continuous outcome, $n_1=1000$, $n_2=400$, $2000$ simulations. Bias, ASE, ESE, SER, RMSE, and 95\% CI coverage calculated for the ATE.}
  \label{tab:simrescontinuous}
\centering
\setlength{\tabcolsep}{3.5pt} 

\begin{tabular}{c l r r r r r c c c} 
\hline
  Scenario & Estimator & Bias & ASE & ESE & SER & RMSE & \makecell{95\% CI \\  Coverage (\%)} & \makecell{Mean \\ Diagnostic \\ Statistic} & \makecell{Diagnostic \\ 95\% CI \\  Includes Zero (\%)}\\ 
  \hline 
1	&	Naïve MS	&	-0.1	&	5.02	&	4.96	&	1.01	&	5.0	&	95	&	0.0	&	94	\\	
	&	Naïve SS	&	-0.1	&	3.55	&	3.51	&	1.01	&	3.5	&	96	&		&		\\	
	&	Bridging MS	&	-0.1	&	5.02	&	4.96	&	1.01	&	5.0	&	95	&	0.0	&	95	\\	
	&	Bridging SS	&	0.0	&	2.16	&	2.16	&	1.00	&	2.2	&	95	&		&		\\	\\
2	&	Naïve MS	&	17.8	&	5.65	&	5.64	&	1.00	&	18.7	&	11	&	20.3	&	0	\\	
	&	Naïve SS	&	38.1	&	4.30	&	4.34	&	0.99	&	38.3	&	0	&		&		\\	
	&	Bridging MS	&	0.0	&	6.01	&	5.93	&	1.01	&	5.9	&	96	&	0.1	&	95	\\	
	&	Bridging SS	&	0.0	&	3.66	&	3.69	&	0.99	&	3.7	&	95	&		&		\\	\\
3	&	Naïve MS	&	37.1	&	5.49	&	5.51	&	1.00	&	37.5	&	0	&	1.0	&	93	\\	
	&	Naïve SS	&	38.1	&	4.30	&	4.34	&	0.99	&	38.3	&	0	&		&		\\	
	&	Bridging MS	&	19.1	&	5.75	&	5.70	&	1.01	&	19.9	&	9	&	-19.0	&	0	\\	
	&	Bridging SS	&	0.0	&	3.66	&	3.69	&	0.99	&	3.7	&	95	&		&		\\	\\
4	&	Naïve MS	&	47.8	&	5.65	&	5.64	&	1.00	&	48.1	&	0	&	20.3	&	0	\\	
	&	Naïve SS	&	68.0	&	4.30	&	4.34	&	0.99	&	68.2	&	0	&		&		\\	
	&	Bridging MS	&	30.0	&	6.01	&	5.93	&	1.01	&	30.5	&	0	&	0.1	&	95	\\	
	&	Bridging SS	&	30.0	&	3.66	&	3.69	&	0.99	&	30.3	&	0	&		&		\\	\\
5	&	Naïve MS	&	57.8	&	5.65	&	5.64	&	1.00	&	58.1	&	0	&	10.3	&	19	\\	
	&	Naïve SS	&	68.1	&	4.30	&	4.34	&	0.99	&	68.2	&	0	&		&		\\	
	&	Bridging MS	&	40.0	&	6.01	&	5.93	&	1.01	&	40.4	&	0	&	-9.9	&	14	\\	
	&	Bridging SS	&	30.0	&	3.66	&	3.69	&	0.99	&	30.3	&	0	&		&		\\	

\hline
\end{tabular}
\begin{tablenotes}
      \item ASE=average estimated standard error; ESE=empirical standard error; SER=standard error ratio (ASE/ESE); RMSE = root mean squared error; CI = confidence interval; SS=single-span; MS=multi-span; Monte Carlo standard error for 95\% CI coverage was 0.5\% when coverage was 95\%.
    \end{tablenotes}
  \end{threeparttable}
\end{table}

\begin{table}
\begin{threeparttable}
\caption{Simulation summary results for binary outcome, $n_1=1000$, $n_2=400$, $2000$ Simulations. Bias, ASE, ESE, SER, RMSE, and 95\% CI coverage calculated for the ATE (formatted as a percent).}
  \label{tab:simresbinary}
\centering
\setlength{\tabcolsep}{3.5pt} 

\begin{tabular}{c l r r r r r c c c} 
\hline
  Scenario & Estimator & Bias & ASE & ESE & SER & RMSE & \makecell{95\% CI \\  Coverage (\%)} & \makecell{Mean \\ Diagnostic \\ Statistic } & \makecell{Diagnostic \\ 95\% CI \\  Includes Zero (\%)}\\ 
  \hline 
1	&	Naïve MS	&	0.1	&	5.39	&	5.48	&	0.98	&	5.5	&	94	&	-0.1	&	94	\\	
	&	Naïve SS	&	0.0	&	3.35	&	3.33	&	1.01	&	3.3	&	95	&		&		\\	
	&	Bridging MS	&	0.0	&	5.40	&	5.49	&	0.98	&	5.5	&	94	&	0.0	&	94	\\	
	&	Bridging SS	&	0.0	&	3.04	&	3.06	&	0.99	&	3.1	&	95	&		&		\\	\\
2	&	Naïve MS	&	-4.5	&	5.46	&	5.39	&	1.01	&	7.0	&	88	&	13.8	&	4	\\	
	&	Naïve SS	&	9.3	&	3.88	&	3.88	&	1.00	&	10.1	&	34	&		&		\\	
	&	Bridging MS	&	-0.2	&	6.47	&	6.46	&	1.00	&	6.5	&	95	&	0.2	&	95	\\	
	&	Bridging SS	&	0.0	&	3.81	&	3.80	&	1.00	&	3.8	&	95	&		&		\\	\\
3	&	Naïve MS	&	11.6	&	5.43	&	5.37	&	1.01	&	12.8	&	43	&	-2.4	&	90	\\	
	&	Naïve SS	&	9.3	&	3.88	&	3.88	&	1.00	&	10.1	&	34	&		&		\\	
	&	Bridging MS	&	19.6	&	6.41	&	6.39	&	1.00	&	20.6	&	14	&	-19.6	&	0	\\	
	&	Bridging SS	&	0.0	&	3.81	&	3.80	&	1.00	&	3.8	&	95	&		&		\\	\\
4	&	Naïve MS	&	6.9	&	5.22	&	5.18	&	1.01	&	8.6	&	74	&	13.8	&	4	\\	
	&	Naïve SS	&	20.7	&	3.54	&	3.56	&	0.99	&	21.0	&	0	&		&		\\	
	&	Bridging MS	&	16.4	&	6.10	&	6.11	&	1.00	&	17.5	&	24	&	0.2	&	95	\\	
	&	Bridging SS	&	16.6	&	3.56	&	3.58	&	0.99	&	16.9	&	1	&		&		\\	\\
5	&	Naïve MS	&	13.2	&	5.31	&	5.29	&	1.00	&	14.2	&	30	&	7.5	&	53	\\	
	&	Naïve SS	&	20.7	&	3.54	&	3.56	&	0.99	&	21.0	&	0	&		&		\\	
	&	Bridging MS	&	25.4	&	6.14	&	6.18	&	0.99	&	26.1	&	2	&	-8.8	&	43	\\	
	&	Bridging SS	&	16.6	&	3.56	&	3.58	&	0.99	&	16.9	&	1	&		&		\\	

\hline
\end{tabular}
\begin{tablenotes}
      \item ASE=average estimated standard error; ESE=empirical standard error; SER=standard error ratio (ASE/ESE); RMSE = root mean squared error; CI = confidence interval; SS=single-span; MS=multi-span; Monte Carlo standard error for 95\% CI coverage was 0.5\% when coverage was 95\%.
    \end{tablenotes}
  \end{threeparttable}
\end{table}

\section{Example}\label{sec:application}
\label{sec:example}
The aim of this analysis is to compare the effect of two-drug (dual) and four-drug (quad) ARV regimens on (1) mean CD4 cell counts and (2) the proportion of persons with CD4 counts > 250 cells/mm$^{3}$ at 8 weeks in persons with advanced HIV. Because of the existence of three-drug (triple) therapy and research showing that three-drug therapy is more efficacious than two-drug therapy, no trials directly comparing two- and four-drug regimens were conducted. Instead, these regimens are compared using the single- and multi-span bridging estimators based on data from ACTG 320 and ACTG 388, two randomized controlled trials conducted to compare the efficacy and safety of ARV regimens in persons with advanced HIV. 

ACTG 320 was a multi-site double-blind trial conducted in the United States. The 1156 participants were randomized to receive dual or triple therapy. Dual therapy consisted of zidovudine (200 mg three times daily), lamivudine (150 mg two times daily), and a placebo. Triple therapy included the same doses of zidovudine and lamivudine but replaced the placebo with indinavir (800 mg three times daily). Participants had to be more than 16 years old and HIV-positive with CD4 count of 200 cells/mm$^{3}$ or less within 60 days prior to baseline. Additionally, participants had at least 3 months of prior zidovudine treatment and no more than one week of prior lamivudine treatment. Those with prior treatment with protease inhibitors were ineligible. Participants were evaluated at baseline, week 4, week 8, and every 8 weeks thereafter. The study was terminated early, after approximately one year of enrollment, due to efficacy of  triple therapy relative to dual therapy. Additional eligibility criteria and implementation details are described elsewhere \citep{hammer1997controlled}. 

The efficacy of triple and quad ARV regimens were compared in ACTG 388, an open label randomized trial conducted in the United States and Italy. The 517 participants were randomized to triple therapy or one of two four-drug therapies. The triple therapy arm received lamivudine (150 mg two times daily), zidovudine (300 mg two times daily), and indinavir (800 mg three times daily). Two different four-drug regimens were considered, both increasing the dose of indinavir to 1000 mg three times daily and adding efavirenz (600 mg, once daily) or nelfinavir (1250 mg, two times daily). Participants had CD4 counts of 200 cells/mm$^{3}$ or lower or plasma HIV-1 RNA levels of 80,000 copies/mL or greater at screening. Participants had no or limited prior ARV therapy; prior ARV therapy was restricted to zidovudine, stavudine, didanosine, and zalcitabine. Participants were evaluated at baseline, week 4, week 8, and every 8 weeks thereafter with a maximum follow-up of 96 weeks. Additional details about the trial are described elsewhere \citep{fischl2003randomized}. 

To harmonize data from ACTG 320 and 388, common levels of covariates were derived. Three participants with missing CD4 at baseline were excluded from the analysis. The outcome was defined as CD4 cell counts at 8 weeks, a common evaluation time in the two studies. Participants who died prior to week 8 (4 participants from ACTG 320 and 1 participant from ACTG 388) had their CD4 cell count at week 8 set to 0. As previously noted, ACTG 320 required at least 3 months of prior zidovudine treatment, but only 31 (9\%) ACTG 388 participants had prior ARV therapy. This sample size did not allow for restricting the ACTG 388 sample by prior HIV treatment regimen, which is a limitation of this analysis. Note only the efavirenz quad treatment is considered in this analysis, as this quad therapy was found superior to triple therapy in ACTG 388. Quad therapy including nelfinavir was estimated to be inferior to triple therapy. 

Baseline characteristics of the two trials are compared in Table \ref{tab:AppTable}. Participants in the two trials had similar marginal distributions by gender, race/ethnicity, history of IDU, Karnofsky score, and age. However, CD4 cell counts differed substantially at baseline, with a higher mean CD4 cell count in ACTG 388 compared to ACTG 320. The distribution in baseline CD4 was expected to vary between trials due to differing inclusion criteria. Participants in ACTG 388 were eligible with either CD4 counts of 200 cells/mm$^{3}$ or lower or plasma HIV-1 RNA levels of 80,000 copies/mL or greater at screening, while ACTG 320 required CD4 counts of 200 cells/mm$^{3}$ or lower.

The single and multi-span bridging estimators were applied to data from ACTG 320 and 388 to compare (1) mean CD4 cell counts and (2) the proportion of persons with CD4 counts > 250 cells/mm$^{3}$ after 8 weeks of treatment between the two- and four-drug (efavirenz) ARV regimens described above. Participants of ACTG 388 were assumed to be representative of the target population, and ACTG 320 participants were representative of a non-focal population. Inverse-odds of sampling weights were estimated from logistic regression models with stacked data from the two trials, regressing the trial indicator on the following covariates measured at baseline: gender (male vs. female), race (Non-Hispanic Black vs. other), ethnicity (Hispanic vs. Non-Hispanic), injection drug use (never vs. current or previous), Karnofsky score category ($< 90$, $< 100$ and $ \geq 90$, or $100$), age, and CD4 cell count. Age and CD4 cell count were modeled using restricted cubic splines with four knots placed at the 5th, 35th, 65th, and 95th percentiles \citep{perperoglou2019review}. Missingness in the outcome variable ranged from 6.4\% in the ACTG 388 quad treatment arm to 15.4\% in the ACTG 320 dual treatment arm. Inverse probability of missingness weights were estimated from separate logistic regression models fit on each study arm, with the same set of covariates included as the transportability model. Karnofsky score category was collapsed for the four-drug therapy arm in ACTG 388. 

\begin{table}
\begin{threeparttable}
\caption{Baseline characteristics of ACTG 320 and ACTG 388 participants by treatment arm}
\label{tab:AppTable}
\centering

\begin{tabular}{l l c c c c} 
\hline
 & & \multicolumn{2}{c}{ACTG 320} & \multicolumn{2}{c}{ACTG 388} \\
 &  & \makecell{Dual \\ $n=579$} & \makecell{Triple \\ $n=577$} & \makecell{Triple \\ $n=168$} & \makecell{Quad (EFZ) \\ $n=173$} \\
\hline
\\
Male	&	&	485 (84\%)	&	471 (82\%)	&	129 (77\%) &	143 (83\%)	\\ \\
Hispanic	&	&	106 (18\%)	&	99 (17\%)	&	29 (17\%) &	35 (20\%)	\\ \\
Non-Hispanic Black	&	&	165 (28\%)	&	163 (28\%)	&	52 (31\%) &	54 (31\%)	\\ \\
History of IDU	&	&	93 (16\%)	&	91 (16\%)	&	21 (13\%) &	25 (14\%)	\\ \\
Karnofsky Score Category	&	< 90	&	108 (19\%)	&	106 (18\%)	&	32 (19\%) &	22 (13\%)	\\
&	$\geq$ 90, < 100	&	269 (46\%)	&	276 (48\%)	&	77 (46\%) &	90 (52\%)	\\
&	100	&	202 (35\%)	&	195 (34\%)	&	59 (35\%) &	61 (35\%)	\\ \\
Age	&	Median (Q1,Q3)	&	38 (33, 44)	&	38 (33, 44)	&	37 (32, 45) &	36 (31, 42)	\\
&	Mean (SD)	&	39 (9)	&	39 (9)	&	39 (10) &	37 (9)	\\
&	Min, Max	&	17, 73	&	16, 74	&	19, 76	& 21, 63	\\ \\
CD4	cells/mm$^{3}$ &	Median (Q1,Q3)	&	70 (23, 135)	&	80 (24, 138)	&	125 (32, 226)	&	96 (32, 199)	\\
&	Mean (SD)	&	85 (70)	&	89 (70)	&	160 (154)	&	147 (151)	\\
&	Min, Max	&	0, 392	&	0, 348	&	0, 728	&	0, 728		\\
& Missing & 1	&	0	&	1	&	1	\\
  \hline 
\end{tabular}
\begin{tablenotes}
      \item Note: EFZ=efavirenz; IDU=Injection Drug Use; SD=Standard Deviation
    \end{tablenotes}
  \end{threeparttable}
\end{table}

Because of observed differences in CD4 cell counts at baseline, the analysis was conducted for the full ACTG samples as well as restricted samples (baseline CD4 $ \leq 400$, $\leq 300$, and $\leq 200$). Box plots for the distribution of each set of weights are presented in Figure \ref{fig:weights} in the Appendix.

\begin{figure} 
\begin{center}
  \includegraphics[width=.75\columnwidth]{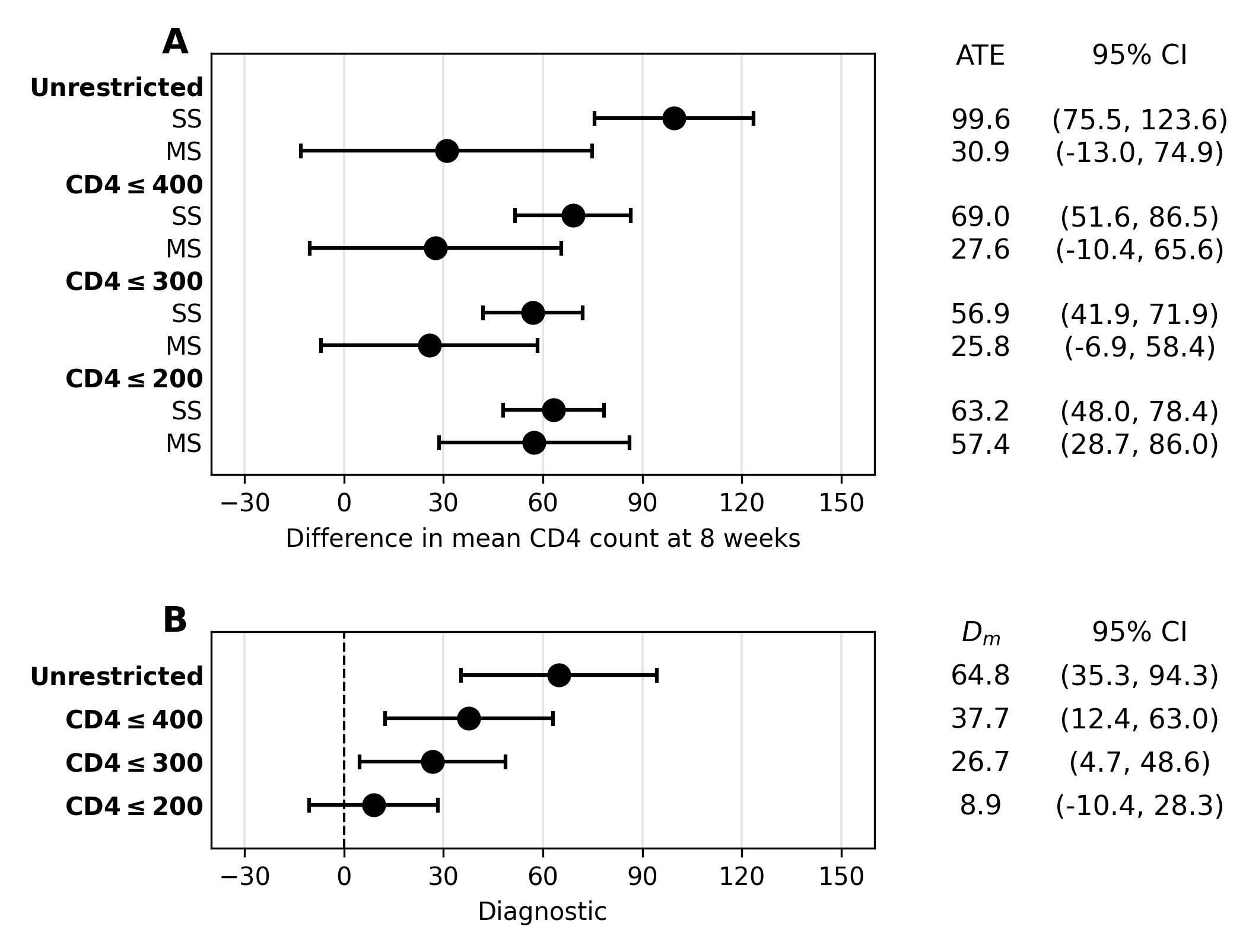}
  \caption{ATE comparing four-drug to two-drug ART in the target (ACTG 388) population, single-span and multi-span estimates with 95\% confidence intervals: (A) mean CD4 cell count and (B) diagnostic plot of estimated difference in the outcome in the shared (three-drug) arm with 95\% confidence intervals, target vs. non-focal population}
    \label{fig:Appres1}
\end{center}
\end{figure}

\begin{figure} 
\begin{center}
  \includegraphics[width=.75\columnwidth]{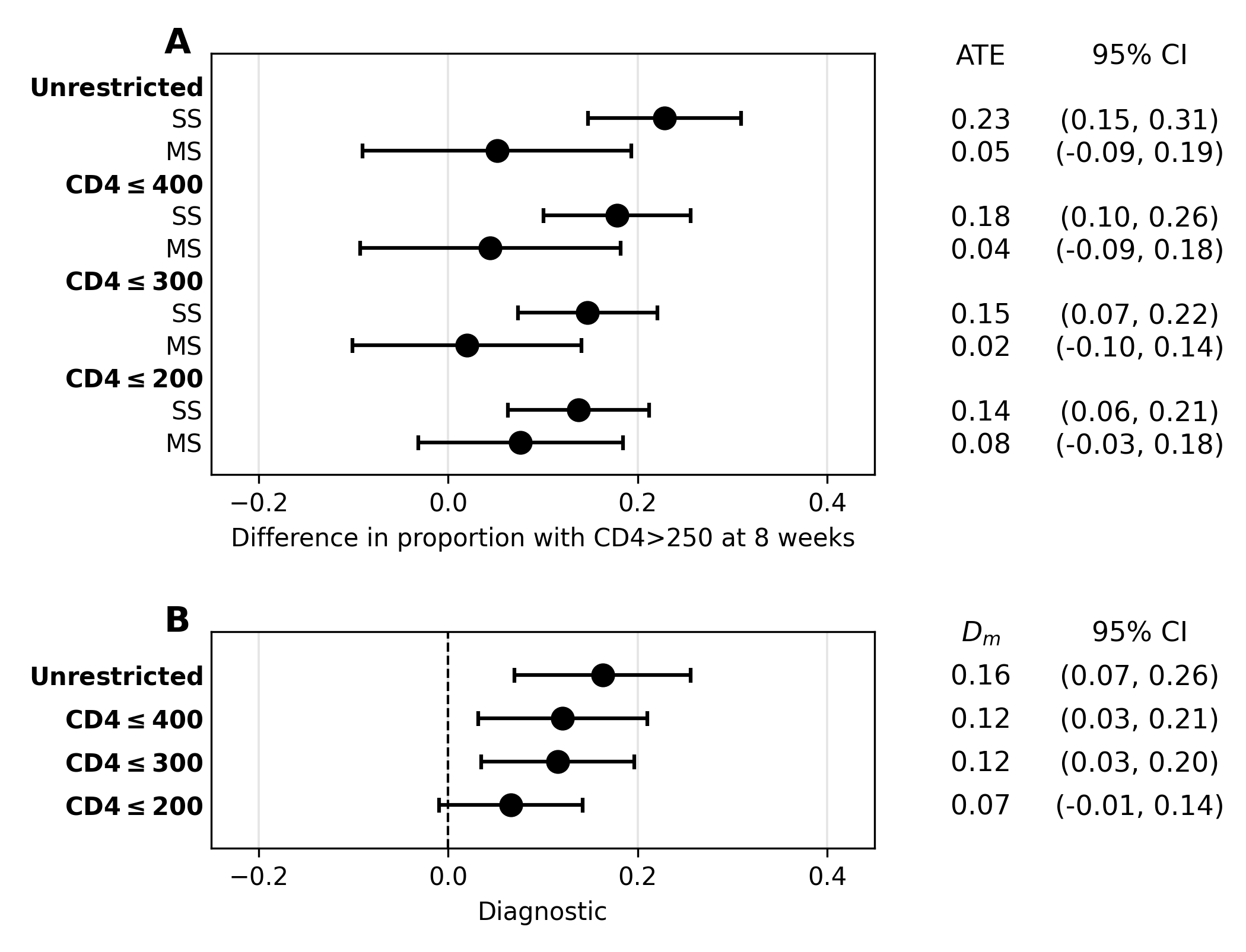}
  \caption{ATE comparing four-drug to two-drug ART in the target (ACTG 388) population, single-span and multi-span estimates with 95\% confidence intervals: (A) proportion with CD4 count > 250 cells/mm$^{3}$ and (B) diagnostic plot of estimated difference in the outcome in the shared (three-drug) arm with 95\% confidence intervals, target vs. non-focal population}
    \label{fig:Appres2}
\end{center}
\end{figure}

Differences in ACTG 320 and 388 CD4 cell count eligibility criteria made the identification assumptions questionable for the unrestricted samples. Without restrictions by baseline CD4 cell count, there were discrepancies in standardized estimates of mean CD4 at 8 weeks in the shared arms. These discrepancies are clear in the diagnostic plots (Figures \ref{fig:Appres1}B and \ref{fig:Appres2}B), where estimated differences in mean CD4 in the shared arms exceed zero and 95\% CIs exclude zero for both the continuous and binary outcomes. The diagnostic provides evidence of a violation of identification assumptions for the multi-span estimator, which resulted in substantive differences between the single- and multi-span estimates of the ATE for the unrestricted samples (Figures \ref{fig:Appres1}A and \ref{fig:Appres2}A). As the samples were increasingly restricted by baseline CD4 cell counts, the estimated single- and multi-span ATEs were increasingly similar for both outcomes (Figures \ref{fig:Appres1}A and \ref{fig:Appres2}A). Restricting by baseline CD4 was also supported by the diagnostic, as the 95\% CIs for the diagnostic included zero for both outcomes when baseline CD4 was restricted to be $\leq 200$. This threshold coincides with the more restrictive CD4 eligibility criteria for ACTG 320. As in the simulation study, the single-span estimator was more precise than the multi-span estimator. The single-span CIs were narrower than multi-span CIs for both outcomes and each level of restriction by baseline CD4 considered. When the data were restricted to participants with baseline CD4 cell counts $\leq 200$, the estimated ATE of four-drug therapy versus two-drug therapy on CD4 cell count at 8 weeks was 63.2 ($95\%$ CI: 48.0, 78.4) for the single-span estimator and 57.4 ($95\%$ CI: 28.7, 86.0) for the multi-span estimator for the continuous outcome. For the binary outcome, the single- and multi-span estimates of the ATE were 0.14 ($95\%$ CI: 0.06, 0.21) and 0.08 ($95\%$ CI: -0.03, 0.18), respectively.

\section{Discussion}
\label{sec:discussion}
It is sometimes of interest to compare a new therapy to a historic therapy or placebo, but typically such comparisons cannot be made directly in a randomized trial due to ethical or resource limitations \citep{temple2000placebo}. Furthermore, direct comparisons of treatments evaluated in different trials can be misleading, as they fail to account for differences in trial populations \citep{fleming2008current}. Comparisons of therapies evaluated in different trials must account for these differences to yield valid estimates. Methods that allow for valid comparisons across trials may be particularly useful in trials where new therapies are evaluated relative to an active control, but efficacy relative to a placebo is of interest \citep{mauri2017challenges}.

Data fusion provides a framework for combining data from multiple studies to answer questions that could not be answered (as well) by any subset of studies \citep{colefusionaje}. The single- and multi-span bridging estimators considered in this paper are fusion estimators that allow for comparisons of treatments evaluated in different trials and thus comparisons could not be made with data from only one of the trials. The single-span estimator requires fewer identification assumptions, as it only requires identification assumptions to hold in trial arms that are directly compared and not in shared or additional trial arms not of interest. This is particularly advantageous in settings where two trials do not share a common arm or where common arms differ in key ways (e.g., different therapy regimens). Furthermore, the single-span estimator was more precise than the multi-span estimator in all simulation scenarios considered and in the application. 

The proposed diagnostic statistic compares the outcome in standardized shared arms and thus evaluates identification assumptions for the multi-span estimator that are not required for the single-span estimator. In practice, the diagnostic can be applied whenever two trials share a common arm, regardless of whether the single-span or multi-span estimator is ultimately applied. When applied for use with the single-span estimator, the diagnostic does not provide direct evidence of violated assumptions. However, in settings where violations of assumptions in the shared arms are likely to coincide with violations of assumptions in the arms of interest, the diagnostic may still be useful. 

The validity of identification assumptions is critical for the application of bridging estimators. Trial eligibility criteria can be harmonized as much as possible and sensitivity analyses conducted to examine discrepancies in trial populations, where feasible. Key differences between trial populations that are either unmeasured or pose violations of the sampling positivity assumption limit the validity of bridged comparisons. However, in cases where trials have similar inclusion and exclusion criteria or share sufficient subsets of participants with comparable inclusion and exclusion covariates, bridged treatment comparisons allow for consistent estimation of treatment effects for therapies evaluated in different trials under the stated identification assumptions. Such estimators may be particularly applicable when trials are specifically designed for comparability, e.g., when master protocols are implemented \citep{woodcock2017master}. 

This work can be extended in a number of areas. While parametric models were used to estimate inverse-probability of missingness weights and inverse odds of sampling weights, these weights could be estimated nonparametrically for low dimensional, discrete covariate sets. Machine learning methods can be used to estimate inverse-probability of missingness weights and inverse odds of sampling weights, which would make estimators more robust to model misspecification. However, there are challenges with the use of machine learning methods for estimation of nuissance model parameters \citep{ZivichML,Chernozhukov}. Augmented inverse probability weighted estimators could also be considered which would be consistent when one of two specified models is correctly specified rather than relying on correct specification of a single model.

\section*{Supplementary Material} 
R and Python code for computing the different estimators along with the corresponding standard error estimators is available at https://github.com/bonnieshook/BridgedTreatmentComparisons.

\section*{Bibliography}

	\bibliographystyle{apalike}
	\bibliography{bibliography}

\newcommand{\noop}[1]{}
\begin{thebibliography}{}

\bibitem[Breskin et~al., 2021]{breskin2021fusion}
Breskin, A., Cole, S.~R., Edwards, J.~K., Brookmeyer, R., Eron, J.~J., and
  Adimora, A.~A. (2021).
\newblock Fusion designs and estimators for treatment effects.
\newblock {\em {Statistics in Medicine}}, 40(13):3124--3137.

\bibitem[Catal{\'a}-L{\'o}pez et~al., 2014]{catala2014transitive}
Catal{\'a}-L{\'o}pez, F., Hutton, B., and Moher, D. (2014).
\newblock The transitive property across randomized controlled trials: if b is
  better than a, and c is better than b, will c be better than a?
\newblock {\em Revista Espanola de Cardiologia (English ed.)}, 67(8):597--602.

\bibitem[Chernozhukov et~al., 2018]{Chernozhukov}
Chernozhukov, V., Chetverikov, D., Demirer, M., Duflo, E., Hansen, C., Newey,
  W., and Robins, J. (2018).
\newblock {Double/debiased machine learning for treatment and structural
  parameters}.
\newblock {\em The Econometrics Journal}, 21(1):C1--C68.

\bibitem[Cole et~al., 2022]{colefusionaje}
Cole, S.~R., Edwards, J.~K., Breskin, A., Rosin, S., Zivich, P.~N., Shook-Sa,
  B.~E., and Hudgens, M.~G. (2022).
\newblock {Illustration of 2 Fusion Designs and Estimators}.
\newblock {\em American Journal of Epidemiology}, 192(3):467--474.

\bibitem[Dahabreh et~al., 2020]{dahabreh2020extending}
Dahabreh, I.~J., Robertson, S.~E., Steingrimsson, J.~A., Stuart, E.~A., and
  Hernan, M.~A. (2020).
\newblock Extending inferences from a randomized trial to a new target
  population.
\newblock {\em Statistics in Medicine}, 39(14):1999--2014.

\bibitem[Donnell, 2022]{Donnell}
Donnell, D. (2022).
\newblock Counterfactual estimation of cab-la efficacy against placebo using
  external trial data.
\newblock Presented at: Conference on Retroviruses and Opportunistic
  Infections; February 12, 2022.

\bibitem[Fischl et~al., 2003]{fischl2003randomized}
Fischl, M.~A., Ribaudo, H.~J., Collier, A.~C., Erice, A., Giuliano, M.,
  Dehlinger, M., Eron~Jr, J.~J., Saag, M.~S., Hammer, S.~M., Vella, S., et~al.
  (2003).
\newblock A randomized trial of 2 different 4-drug antiretroviral regimens
  versus a 3-drug regimen, in advanced human immunodeficiency virus disease.
\newblock {\em The Journal of Infectious Diseases}, 188(5):625--634.

\bibitem[Fleming, 2008]{fleming2008current}
Fleming, T.~R. (2008).
\newblock Current issues in non-inferiority trials.
\newblock {\em Statistics in Medicine}, 27(3):317--332.

\bibitem[Glidden et~al., 2020]{glidden2020bayesian}
Glidden, D.~V., Stirrup, O.~T., and Dunn, D.~T. (2020).
\newblock A bayesian averted infection framework for prep trials with low
  numbers of hiv infections: application to the results of the discover trial.
\newblock {\em The Lancet HIV}, 7(11):e791--e796.

\bibitem[Hammer et~al., 1997]{hammer1997controlled}
Hammer, S.~M., Squires, K.~E., Hughes, M.~D., Grimes, J.~M., Demeter, L.~M.,
  Currier, J.~S., Eron~Jr, J.~J., Feinberg, J.~E., Balfour~Jr, H.~H., Deyton,
  L.~R., et~al. (1997).
\newblock A controlled trial of two nucleoside analogues plus indinavir in
  persons with human immunodeficiency virus infection and cd4 cell counts of
  200 per cubic millimeter or less.
\newblock {\em New England Journal of Medicine}, 337(11):725--733.

\bibitem[Hughes, 2020]{Hughes}
Hughes, J. (2020).
\newblock Developing placebo counterfactuals for prep studies.
\newblock Available at:
  \url{https://www.hptn.org/sites/default/files/inline-files/11_HPTN\%202020\%20Update\%20Webinar_Stats.pdf}.

\bibitem[Lumley, 2002]{lumley2002network}
Lumley, T. (2002).
\newblock Network meta-analysis for indirect treatment comparisons.
\newblock {\em Statistics in Medicine}, 21(16):2313--2324.

\bibitem[Mauri and D’Agostino~Sr, 2017]{mauri2017challenges}
Mauri, L. and D’Agostino~Sr, R.~B. (2017).
\newblock Challenges in the design and interpretation of noninferiority trials.
\newblock {\em New England Journal of Medicine}, 377(14):1357--1367.

\bibitem[Perperoglou et~al., 2019]{perperoglou2019review}
Perperoglou, A., Sauerbrei, W., Abrahamowicz, M., and Schmid, M. (2019).
\newblock A review of spline function procedures in r.
\newblock {\em BMC Medical Research Methodology}, 19(1):1--16.

\bibitem[Saul and Hudgens, 2020]{saul2020calculus}
Saul, B.~C. and Hudgens, M.~G. (2020).
\newblock The calculus of m-estimation in r with geex.
\newblock {\em Journal of Statistical Software}, 92(2).

\bibitem[Stefanski and Boos, 2002]{stefanski2002calculus}
Stefanski, L.~A. and Boos, D.~D. (2002).
\newblock The calculus of m-estimation.
\newblock {\em The American Statistician}, 56(1):29--38.

\bibitem[Temple and Ellenberg, 2000]{temple2000placebo}
Temple, R. and Ellenberg, S.~S. (2000).
\newblock Placebo-controlled trials and active-control trials in the evaluation
  of new treatments. part 1: ethical and scientific issues.
\newblock {\em Annals of Internal Medicine}, 133(6):455--463.

\bibitem[Westreich et~al., 2017]{westreich2017transportability}
Westreich, D., Edwards, J.~K., Lesko, C.~R., Stuart, E., and Cole, S.~R.
  (2017).
\newblock Transportability of trial results using inverse odds of sampling
  weights.
\newblock {\em American Journal of Epidemiology}, 186(8):1010--1014.

\bibitem[Woodcock and LaVange, 2017]{woodcock2017master}
Woodcock, J. and LaVange, L.~M. (2017).
\newblock Master protocols to study multiple therapies, multiple diseases, or
  both.
\newblock {\em New England Journal of Medicine}, 377(1):62--70.

\bibitem[Yuan and Jennrich, 1998]{yuan1998asymptotics}
Yuan, K.-H. and Jennrich, R.~I. (1998).
\newblock Asymptotics of estimating equations under natural conditions.
\newblock {\em Journal of Multivariate Analysis}, 65(2):245--260.

\bibitem[Zhang et~al., 2016]{zhang2016new}
Zhang, Z., Nie, L., Soon, G., and Hu, Z. (2016).
\newblock New methods for treatment effect calibration, with applications to
  non-inferiority trials.
\newblock {\em Biometrics}, 72(1):20--29.

\bibitem[Zivich et~al., 2022a]{ZivichML}
Zivich, P.~N., Breskin, A., and Kennedy, E.~H. (2022a).
\newblock Machine learning and causal inference.
\newblock {\em Wiley StatsRef: Statistics Reference Online}.

\bibitem[Zivich et~al., 2022b]{zivich2022bridged}
Zivich, P.~N., Cole, S.~R., Edwards, J.~K., Shook-Sa, B.~E., Breskin, A., and
  Hudgens, M.~G. (2022b).
\newblock Bridged treatment comparisons: an illustrative application in hiv
  treatment.
\newblock {\em arXiv preprint: 2206.04445}.

\bibitem[Zivich et~al., 2022c]{zivich2022delicatessen}
Zivich, P.~N., Klose, M., Cole, S.~R., Edwards, J.~K., and Shook-Sa, B.~E.
  (2022c).
\newblock Delicatessen: M-estimation in python.
\newblock {\em arXiv preprint: 2203.11300}.

\end{thebibliography}

\section*{Acknowledgements} 
This research was funded in part through Developmental funding from the University of North Carolina at Chapel Hill Center for AIDS Research (CFAR), an NIH funded program P30 AI050410. This work was also supported by US National Institutes of Health under award numbers R01 AI157758 and R01 AI085073. The content is solely the responsibility of the authors and does not necessarily represent the official views of the National Institutes of Health. We thank Dr. Michael Hughes, Principal Investigator for the Statistical and Data Management Center for the AIDS Clinical Trial Group, for providing public use study data and Mingwei Fei of the University of North Carolina at Chapel Hill for helpful feedback. \vspace*{-8pt}

	\clearpage
	
	\begin{appendices}
		\setcounter{equation}{0}
		\renewcommand{\theequation}{A.\arabic{equation}}
\section*{Appendix}

    \subsection*{Identification Proofs (Section \ref{sec:methods-Idassumptions})}

The ITT parameter ${ATE}^{3-1}$ can be expressed in a single-span form, i.e., ${ATE}^{3-1}=E(Y^3 \mid R=2)-E(Y^1 \mid R=2)$, or a multi-span form, i.e., $ATE^{3-1}=E(Y^3 \mid R=2)-E(Y^2 \mid R=2)+E(Y^2 \mid R=2)-E(Y^1 \mid R=2)$. Both forms of $ATE^{3-1}$ are identifiable under the assumptions in Table \ref{tab:identcond}. Recall that $\pi_A(a,r)=Pr(A=a \mid R=r, X)$, $\pi_M(a,r)=Pr(M=0 \mid A=a, R=r, X)$, and $\pi_R=Pr(R=1 \mid X)$. Then, for $a \in \{2, 3\}$
\begin{align}
E(Y^a \mid R=2)&=E\left\{ \frac{I(A=a)Y^a}{\pi_A(a,2)} \bigg| R=2  \right \} \label{id1step1} \\
&=E\left[ E\left\{ \frac{I(A=a)Y^a}{\pi_A(a,2)} \bigg| R=2,A,X  \right \} \bigg| R=2 \right] \label{id1step2} \\
&=E\left[ \frac{I(A=a)}{\pi_A(a,2)}E\left\{ \frac{I(M=0)Y^a}{\pi_M(a,2)} \bigg| R=2,A,X  \right \} \bigg| R=2 \right] \label{id1step3} \\
&=E\left\{ \frac{I(A=a)I(M=0)Y}{\pi_A(a,2)\pi_M(a,2)} \bigg| R=2 \right \}  \label{id1step4} \\
&=Pr(R=2)^{-1}E\left\{ \frac{I(A=a)I(M=0)I(R=2)Y}{\pi_A(a,2)\pi_M(a,2)} \right \} \label{id1step5}
  \end{align}
where (\ref{id1step1}) holds by assumptions 5 and 6 in Table \ref{tab:identcond}, (\ref{id1step2}) by iterated expectation, (\ref{id1step3}) by assumptions 3 and 4, (\ref{id1step4}) by assumption 2 and iterated expectation, and (\ref{id1step5}) by definition of conditional expectation. All steps require assumption 1 to hold. Note (\ref{id1step5}) is identifiable from observed data from the target population.  

Likewise, for $a \in \{1, 2\}$ and following an approach similar to \cite{dahabreh2020extending}, \begin{align}
		E(Y^a \mid R = 2)\Pr(R=2) &= E\left(Y^a \mid R=2\right)\Pr(R=2)  \label{id2step1} \\ 
		&= E\left\{I(R=2) Y^a \right\} \label{id2step2} \\
		&= E\left[E\left\{I(R=2)Y^a \mid X \right\} \right] \label{id2step3} \\
		&= E\left[E\left\{I(R=2) \mid X \right\} E\left(Y^a \mid X \right)\right] \label{id2step4} \\
		&= E\left\{(1-\pi_R)E(Y^a \mid X)\right\} \label{id2step5} \\
		&= E\left[\frac{1-\pi_R}{\pi_R}
		E\left\{Y^a I(R=1) \mid X\right\}		
		\right] \label{id2step6} \\
		&=E\left\{\frac{1-\pi_R}{\pi_R}  I(R=1) Y^a \right\} \label{id2step7} \\		
		&=E\left[E\left\{\frac{1-\pi_R}{\pi_R} Y^a I(R=1) \bigg | X, R \right\} \right] \label{id2step8} \\
        &=E\left[\frac{1-\pi_R}{\pi_R} I(R=1)E\left\{\frac{I(A=a)}{\pi_A(a,1)} Y^a \bigg | X, R \right\} \right] \label{id2step9} \\
		&=E\left[E\left\{\frac{1-\pi_R}{\pi_R \pi_A(a,1)} I(R=1)I(A=a)Y^a \bigg | X, R,A \right\} \right] \label{id2step10} \\
		&=E\left[\frac{1-\pi_R}{\pi_R \pi_A(a,1)} I(R=1)I(A=a)E\left\{\frac{I(M=0)}{\pi_M(a,1)}Y^a \bigg | X, R,A \right\} \right] \label{id2step11} \\
		&=E\left\{\frac{1-\pi_R}{\pi_R \pi_A(a,1) \pi_M(a,1)} I(R=1)I(A=a)I(M=0)Y^a  \right\} \label{id2step12} \\
		&=E\left\{\frac{1-\pi_R}{\pi_R \pi_A(a,1) \pi_M(a,1)} I(R=1)I(A=a)I(M=0)Y  \right\}	\label{id2step13} \end{align} where (\ref{id2step1}) holds by multiplication by 1, (\ref{id2step2}) by definition of conditional expectation, (\ref{id2step3}) by iterated expectation, (\ref{id2step4}) by assumptions 7 and 8 in Table~\ref{tab:identcond}, (\ref{id2step5}) by definition of $\pi_R$, (\ref{id2step6}) by assumption 7, (\ref{id2step7}) and (\ref{id2step8}) by iterated expectation, (\ref{id2step9}) by assumptions 5 and 6, (\ref{id2step10}) by iterated expectation, (\ref{id2step11}) by assumptions 3 and 4, (\ref{id2step12}) by iterated expectation, and (\ref{id2step13}) by assumption 2. All steps require assumption 1 to hold. Note (\ref{id2step1}) is identifiable from observed data from the non-focal population.

  As noted in Section \ref{sec:methods-Idassumptions}, the single-span and multi-span versions of ${ATE}^{3-1}$ are then identified by contrasting each of their component parts, with $E(Y^3 \mid R=2)$ and $E(Y^1 \mid R=2)$ identified using data from the target and non-focal populations, respectively. In the multi-span form, $E(Y^2 \mid R=2)$ is identified separately using data from each of the trials. 

\subsection*{Motivation of Hajek Estimators (Section \ref{sec:methods-estimators})}
  The Hajek estimator $\hat{E}_1(Y^a \mid R=2;\hat{\eta}_e)$ is motivated from the identified form of the causal mean (\ref{eq:idfinnfp}) by noting  (following an approach similar to \cite{dahabreh2020extending})  \begin{align}
		&E\left\{\frac{(1-\pi_R)I(A=a)I(R=1)I(M=0)}{\pi_R\pi_A(a,1)\pi_M(a,1)}\right\} \notag \\
        &=	E\left\{\frac{(1-\pi_R)I(A=a, R=1, M=0)}{\Pr(M=0 \mid A =a, R = 1, X)\Pr(A = a \mid R = 1, X)\Pr(R = 1 \mid X)}\right\}  \notag \\
		&=
		E\left\{\frac{(1-\pi_R)I(A=a, R=1, M=0)}{\Pr(M=0, A=a, R=1 \mid X)}\right\}  \notag \\
		&=
		E\left[E\left\{\frac{(1-\pi_R)I(A=a,R=1,M=0)}{\Pr(M=0, A=a, R=1 \mid X)} \bigg | X \right\}  \right] \notag \\
		&=E\left\{\frac{(1-\pi_R)E\left\{I(A=a, R=1, M=0) \mid X\right\}}{\Pr(M=0, A=a, R=1 \mid X)}\right\}  \notag \\
		&= E\left\{\frac{(1-\pi_R)\Pr(M=0 \mid A=a, R=1, X)\Pr(A=a \mid R=1, X)\Pr(R=1 \mid X)}{\pi_R\pi_A(a,1)\pi_M(a,1)}\right\}  \notag \\
		&= E(1 - \pi_R) \notag \\
		&= E\{\Pr(R=2 \mid X)\} \notag   \\
		&= E\left[E\left\{I(R=2) \mid X \right\} \right] \notag \\
		&=\Pr(R=2 )	 \notag \label{eq:idfin_denom} \end{align} 
 
 Similarly, the Hajek estimator $\hat{E}_2(Y^a \mid R=2;\hat{\eta}_e)$ is motivated from the identified form of the causal mean (\ref{eq:idfintp}).
    
	\subsection*{Consistency and Asymptotic Normality of (\ref{est:MS}) and (\ref{est:SS}) (Section \ref{sec:methods-estimators})}

	Let $\theta_{ms} = \{\eta_1, \alpha_2^2, \alpha_2^3, \alpha_1^1, \alpha_1^2, ATE^{3-1}\}$ and $\theta_{ss}=\{\eta_2, \alpha_2^3, \alpha_1^1, ATE^{3-1}\}$ represent the parameter vectors for the multi-span and single-span estimators, respectively, where $\alpha_r^{a}=E_{r}[Y^a \mid R = 2; \eta_e]$ for $e \in \{1, 2\}$ and for $(a, r) \in \{(1, 1), (2, 1), (2, 2), (3, 2)\}$. Note that $\alpha_2^2 = \alpha^2_1$ in the multi-span parameter vector. Recall that in trial 1 there are $n_1$ iid copies of $O_{i} = \{R_i = 1, A_i, M_i, I(M_i = 0)Y_i, X_i\}$, and in trial 2 there are $n_2$ iid copies of $O_{i} = \{R_i = 2, A_i, M_i, I(M_i = 0)Y_i, X_i\}$. Data from the $n_1 + n_2$ participants in the two trials are stacked. Observations from participant $i$ in trial 1 and participant $j$ in trial two are assumed to be independent but not necessarily identically distributed. Assume that $\lim_{n \rightarrow \infty}n_r/n=\pi_r$ for unknown sampling fractions $\pi_r>0$ with $r \in \{1,2\}$ and $\pi_1+\pi_2=1$.

	\paragraph{Multi-span estimator} 
	
	Consider the set of estimating equations for the multi-span estimator: 
	
	\begin{equation} \label{eq:MSEE}
	   \sum_{i=1}^n \Psi_i(O_i; \theta_{ms}) = \begin{bmatrix}
		\sum_{i=1}^n \Psi_1(O_i; \lambda_1, \lambda_2, \lambda_3, \lambda_4)\\
		\sum_{i=1}^n \Psi_2(O_i; \gamma_1) \\
		\sum_{i=1}^n \Psi_3(O_{i}; {\eta}_1,\alpha_2^2)\\
		\sum_{i=1}^n \Psi_4(O_{i}; {\eta}_1,\alpha_2^3)\\
		\sum_{i=1}^n \Psi_5(O_i; {\eta}_1, \alpha_1^1) \\          
		\sum_{i=1}^n \Psi_6(O_i; {\eta}_1, \alpha_1^2)  \\
		\alpha_2^3 - \alpha_2^2 + \alpha_1^2 - \alpha_1^1 - ATE^{3-1}
	\end{bmatrix} = 0. \end{equation}
	
	The vector $\Psi_1$ consists of four sets of estimating equations $\Psi_{1q}$ corresponding to the score equations from the missingness models for arms $q \in \{1, 2, 3, 4\}$ as defined in the main text. Assume the four missingness models have the form $\operatorname{logit}\{P(M_i = 0 \mid X_i; \lambda_q)\} = g_m(X_i)\lambda_q$ for arm $q$ with $q \in \{1, 2, 3, 4\}$. Assume $g_m(X_i)$ includes functions of covariates such that each element of $g_m(X_i)$ has finite range. The vector $\Psi_2$ contains the score equations from the sampling model, fit using data from all four arms. Assume the sampling model has the form $\operatorname{logit}\{P(R_i = 2 \mid X_i; \gamma_1)\} = g_s(X_i)\gamma_1$. Under the assumption that the missingness and sampling models are correctly specified, $\Psi_1$ and $\Psi_2$ are unbiased based on maximum likelihood theory, with solutions $\hat \lambda_q$ for $q \in \{1, 2, 3, 4\}$ and $\hat \gamma_1$, respectively. 
	
	Consider $\Psi_3$ and $\Psi_4$, the estimating equations for $\alpha_2^a$ for $a \in \{2,3\}$. These estimating equations can be specified as $\Psi_j= \sum_{i=1}^n  \hat{W}_{Mi}W_{Ai}(Y_i-\alpha_2^a)I(A_i=a)I(R_i=2)I(M_i=0)=0$ for $j \in \{3,4\}$ and $a \in \{2,3\}$, respectively. Dropping subscripts $i$ for notational ease and treating the weights as known, 
	\begin{align}
		&E\{W_{M}W_{A}(Y-\alpha_2^a)I(A=a)I(R=2)I(M=0)\} \notag \\	
		&=E\{W_{M}W_{A}(Y^a-\alpha_2^a)I(A=a)I(R=2)I(M=0)\} \label{con1step1}\\
		&=E[W_{M}W_{A}I(A=a)I(R=2)E\{(Y^a-\alpha_2^a)I(M=0) \mid A, R, X\}] \label{con1step2} \\
		&=E[W_{M}W_{A}I(A=a)I(R=2)E\{(Y^a-\alpha_2^a) \mid A, R, X\}E\{I(M=0) \mid A, R, X\}] \label{con1step3}\\
		&=E\{W_{A}I(A=a)I(R=2)(Y^a-\alpha_2^a)\} \label{con1step4}\\
		&=E\left[W_{A}I(R=2)E\{I(A=a)(Y^a-\alpha_2^a)\mid X, R=2\}\right]Pr(R=2) \label{con1step5}\\
		&=E\left[W_{A}I(R=2)E\{I(A=a) \mid X, R = 2\}E(Y^a-\alpha_2^a \mid X, R = 2) \right]Pr(R=2) \label{con1step6}\\
		&= E[I(R=2)E\{(Y^a \mid R = 2) - \alpha_2^a\}]Pr(R=2) \label{con1step7}\\
		&= 0 \label{con1step8}
		\end{align}
	where (\ref{con1step1}) holds by causal consistency, (\ref{con1step2}) by iterated expectation, (\ref{con1step3}) because $Y^a \perp M \mid \{A, R, X\}$ for $a \in \{2, 3\}$, (\ref{con1step4}) holds by definition of $W_M$, (\ref{con1step5}) holds by iterated expectation and the law of total expectation, (\ref{con1step6}) because $Y^a \perp A \mid\{X, R = 2\}$, (\ref{con1step7}) holds by definition of $W_A$, and (\ref{con1step8}) holds by definition of $\alpha_2^a$. So $\Psi_3$ and $\Psi_4$ are unbiased estimating equations with solutions $\hat{\alpha}_2^2$ and $\hat{\alpha}_2^3$, respectively.
	
	Consider $\Psi_5$ and $\Psi_6$, the estimating equations for estimating $\alpha_1^a$ for $a \in \{1,2\}$ based on data from the non-focal population. These estimating equations can be written as:
	\[\Psi_j= \sum_{i=1}^n  \hat{W}_{Mi}\hat{W}_{Si}W_{Ai}(Y_i-\alpha_1^a)I(A_i=a)I(R_i=1)I(M_i=0)=0\] for $j \in \{5,6\}$ and $a \in \{1,2\}$, respectively. Dropping subscripts $i$ for notational ease and treating the weights as known, 
	\begin{align}
		&E\{W_{M}W_{S}W_{A}(Y-\alpha_1^a)I(A=a)I(R=1)I(M=0)\} \notag\\
		&=E\{W_{M}W_{S}W_{A}(Y^a-\alpha_1^a)I(A=a)I(R=1)I(M=0)\} \label{con2step1}\\
		&=E[W_{M}W_{S}W_{A}I(A=a)I(R=1)E\{(Y^a-\alpha_1^a)I(M=0) \mid A, R, X\}] \label{con2step2}\\ 
		&= E[W_MW_SW_AI(A=a)I(R=1)E\{(Y^a - \alpha_1^a) \mid A, R, X\}E\{I(M=0)\mid A, R, X\}] \label{con2step3}\\
		&=E\{W_SW_AI(A=a)I(R=1)(Y^a - \alpha_1^a)\} \label{con2step4}\\
		&=E[W_SW_AI(R=1)E\{I(A=a)(Y^a - \alpha_1^a) \mid R,X\}] \label{con2step5}\\		
		&=E[W_SW_AI(R=1)E\{I(A=a)\mid R,X\}E\{(Y^a - \alpha_1^a) \mid R,X\}] \label{con2step6}\\		
		&=E\{W_SI(R=1)(Y^a - \alpha_1^a)\} \label{con2step7}\\
		&=E[E\{W_SI(R=1)(Y^a - \alpha_1^a) \mid X, R=1\}Pr(R=1 \mid X)] \label{con2step8}\\
		&=E[W_SE\{I(R=1)\mid X, R=1\} E\{(Y^a - \alpha_1^a) \mid X, R=2\}Pr(R=1 \mid X)] \label{con2step9}\\
		&=E[Pr(R=2 \mid X)E\{I(R=1)\mid X, R=1\}E\{(Y^a - \alpha_1^a) \mid X, R=2\}] \label{con2step10}\\
		&=E[Pr(R=2 \mid X)E\{I(R=1)\mid R=1\}E\{(Y^a - \alpha_1^a) \mid R=2\}] \label{con2step11}\\  
  	&= 0 \label{con2step12}
		\end{align}
	where (\ref{con2step1}) holds by causal consistency, (\ref{con2step2}) holds by iterated expectation, (\ref{con2step3}) holds because $Y^a \perp M \mid \{A, R, X\}$, (\ref{con2step4}) holds by definition of $W_M$, (\ref{con2step5}) holds by iterated expectation, (\ref{con2step6}) holds because $Y^a \perp A \mid X, R$ for $a \in \{1, 2\}$, (\ref{con2step7}) by definition of $W_A$ and iterated expectation, (\ref{con2step8}) holds by iterated expectation and the law of total expectation, (\ref{con2step9}) holds because $Y^a \perp R \mid X$, (\ref{con2step10}) by cancellation of $Pr(R=1 \mid X)$ with the denominator of $W_S$, (\ref{con2step11}) by iterated expectation, and (\ref{con2step12}) by definition of $\alpha_1^a$. So $\Psi_5$ and $\Psi_6$ are unbiased estimating equations, with solutions $\hat{\alpha}_1^1$ and $\hat{\alpha}_1^2$. 
	
The estimating equation $\Psi_7 = \alpha_2^3-\alpha_2^2 + \alpha_1^2 - \alpha_1^1 - ATE^{3-1}$ is equal to zero, and is thus unbiased, by definition of $ATE^{3-1}$. Because (\ref{est:MS}) is the solution to an unbiased estimating equation vector, it follows under suitable regularity conditions that (\ref{est:MS}) is a consistent and asymptotically normal estimator of $\theta_{ms}$, and its asymptotic variance can be consistently estimated by the empirical sandwich variance estimator \citep{yuan1998asymptotics}. 
	
	\paragraph{Single-span estimator} 
	
	Consider the set of estimating equations for the single-span estimator:
	
	\[\sum_{i=1}^n \Psi_i(O_i; \theta_{ss}) = \begin{bmatrix}
		\sum_{i=1}^n \Psi_1(O_i; \lambda_1, \lambda_4)\\
		\sum_{i=1}^n \Psi_2(O_i; \gamma_2) \\
		\sum_{i=1}^n \Psi_4(O_{i}; {\eta}_2,\alpha_2^3)\\
		\sum_{i=1}^n \Psi_5(O_i; {\eta}_2, \alpha_1^1) \\
		\alpha_2^3 - \alpha_1^1 - ATE^{3-1}
	\end{bmatrix} = 0 \]
	For the single-span estimator, $\Psi_1$ has two components, equivalent to $\psi_{14}$ and $\psi_{11}$ from the multi-span estimator. The vector $\Psi_2$ contains the score equations from the sampling model, fit using data only from arms $q \in \{1, 4\}$. Assume the sampling model has the form $\operatorname{logit}\{P(R_i = 2 \mid X_i; \gamma_2)\} = g_S(X_i)\gamma_2$. The equations $\Psi_4$ and $\Psi_5$ are the same as those from the multi-span estimator, but replace $\eta_1$ with $\eta_2$. Unbiasedness of these estimating equations follows from the proof for the multi-span estimator, noting that $\alpha_2^3 - \alpha_1^1 - ATE^{3-1}$ is equal to zero and is thus unbiased by definition of $ATE^{3-1}$. Note this set of estimating equations does not include estimators for the shared arm, and thus the subset of identification assumptions listed in Section \ref{sec:methods-Idassumptions} for the single-span approach is sufficient to show unbiasedness. Because (\ref{est:SS}) is the solution to an unbiased estimating equation vector, it is a consistent and asymptotically normal estimator of $\theta_{ss}$ under suitable regularity conditions, and the empirical sandwich variance estimator is a consistent estimator for the asymptotic variance of (\ref{est:SS}). 
	
 \subsection*{Consistency and Asymptotic Normality of diagnostic statistic $\widehat{D}_m$ (Section \ref{sec:methods-estimators})}
	
	Recall the estimator for the diagnostic $D_m$ from \ref{est:diagnostic} is $\widehat{D}_{m}=\hat{E}_2(Y^2 \mid R=2;\eta_1)-\hat{E}_1(Y^2 \mid R=2;\eta_1) = \hat \alpha_2^2 - \hat\alpha_1^2$.
	Per the consistency proof above, $\hat \alpha_2^2 \overset{p}{\to}\alpha_2^2$ and $\hat \alpha_1^2\overset{p}{\to}\alpha_1^2$. Because $\alpha_2^2 = \alpha_1^2 = E(Y^2 \mid R = 2)$, $\hat D_m \overset{p}{\to} 0$. Furthermore, $\widehat{D}_m$ can be expressed as the solution to an estimating equation $\Psi_8 = \alpha_2^2 - \alpha_1^2 - D_m$. This estimating equation equals zero when the identification assumptions hold, and is thus unbiased. The equation $\Psi_8$ can be stacked with (\ref{eq:MSEE}). Because $\widehat{D}_m$ is the solution to an unbiased estimating equation vector, it is consistent and asymptotically normal, and its variance can be consistently estimated by the empirical sandwich variance estimator.
	
\subsection*{Supplementary Tables and Figures}

\setcounter{table}{0}
\renewcommand{\thetable}{A\arabic{table}}

\setcounter{figure}{0}
\renewcommand{\thefigure}{A\arabic{figure}}

\begin{table}
\begin{threeparttable}
\caption{Simulation summary results for continuous outcome, $n_1=400$, $n_2=1000$, $2000$ simulations. Bias, ASE, ESE, SER, RMSE, and 95\% CI coverage calculated for the ATE.}
\label{tab:appn1400cont}
\centering
\setlength{\tabcolsep}{3.5pt} 

\begin{tabular}{c l r r r r r c c c} 
\hline
  Scenario & Estimator & Bias & ASE & ESE & SER & RMSE & \makecell{95\% CI \\  Coverage (\%)} & \makecell{Mean \\ Diagnostic} & \makecell{Diagnostic \\ 95\% CI \\  Includes Zero (\%)}\\ 
  \hline 
1	&	Naïve MS	&	0.0	&	5.02	&	5.04	&	1.00	&	5.0	&	94	&	0.1	&	95	\\	
	&	Naïve SS	&	0.1	&	3.55	&	3.49	&	1.02	&	3.5	&	95	&		&		\\	
	&	Bridging MS	&	0.0	&	5.02	&	5.04	&	1.00	&	5.0	&	95	&	0.0	&	95	\\	
	&	Bridging SS	&	0.0	&	2.20	&	2.14	&	1.02	&	2.1	&	96	&		&		\\	\\
2	&	Naïve MS	&	18.0	&	6.33	&	6.37	&	0.99	&	19.1	&	18	&	20.5	&	0	\\	
	&	Naïve SS	&	38.4	&	5.10	&	5.14	&	0.99	&	38.8	&	0	&		&		\\	
	&	Bridging MS	&	0.1	&	6.56	&	6.59	&	1.00	&	6.6	&	95	&	0.2	&	95	\\	
	&	Bridging SS	&	0.2	&	3.63	&	3.65	&	1.00	&	3.7	&	94	&		&		\\	\\
3	&	Naïve MS	&	37.3	&	6.27	&	6.31	&	0.99	&	37.8	&	0	&	1.1	&	94	\\	
	&	Naïve SS	&	38.4	&	5.10	&	5.14	&	0.99	&	38.8	&	0	&		&		\\	
	&	Bridging MS	&	19.2	&	6.47	&	6.51	&	0.99	&	20.3	&	16	&	-18.9	&	0	\\	
	&	Bridging SS	&	0.3	&	3.63	&	3.65	&	1.00	&	3.7	&	95	&		&		\\	\\
4	&	Naïve MS	&	47.9	&	6.32	&	6.36	&	0.99	&	48.4	&	0	&	20.5	&	0	\\	
	&	Naïve SS	&	68.4	&	5.09	&	5.13	&	0.99	&	68.6	&	0	&		&		\\	
	&	Bridging MS	&	30.1	&	6.56	&	6.59	&	1.00	&	30.8	&	0	&	0.2	&	95	\\	
	&	Bridging SS	&	30.2	&	3.63	&	3.65	&	1.00	&	30.5	&	0	&		&		\\	\\
5	&	Naïve MS	&	58.0	&	6.32	&	6.36	&	0.99	&	58.3	&	0	&	10.5	&	20	\\	
	&	Naïve SS	&	68.4	&	5.09	&	5.13	&	0.99	&	68.6	&	0	&		&		\\	
	&	Bridging MS	&	40.1	&	6.56	&	6.59	&	1.00	&	40.7	&	0	&	-9.8	&	25	\\	
	&	Bridging SS	&	30.3	&	3.63	&	3.65	&	1.00	&	30.5	&	0	&		&		\\	

\hline
\end{tabular}
\begin{tablenotes}
      \item ASE=average estimated standard error; ESE=empirical standard error; SER=standard error ratio (ASE/ESE); RMSE = root mean squared error; CI = confidence interval; SS=single-span; MS=multi-span; Monte Carlo standard error for 95\% CI coverage was 0.5\% when coverage was 95\%.
    \end{tablenotes}
  \end{threeparttable}
\end{table}

\begin{table}
\begin{threeparttable}
\caption{Simulation summary results for binary outcome, $n_1=400$, $n_2=1000$, $2000$ simulations. Bias, ASE, ESE, SER, RMSE, and 95\% CI coverage calculated for the ATE (formatted as a percent).}
\label{tab:appn140bin}
\centering
\setlength{\tabcolsep}{3.5pt} 

\begin{tabular}{c l r r r r r c c c} 
\hline
  Scenario & Estimator & Bias & ASE & ESE & SER & RMSE & \makecell{95\% CI \\  Coverage (\%)} & \makecell{Mean \\ Diagnostic } & \makecell{Diagnostic \\ 95\% CI \\  Includes Zero (\%)}\\ 
  \hline 
1	&	Naïve MS	&	0.0	&	5.83	&	5.89	&	0.99	&	5.9	&	95	&	0.1	&	95	\\	
	&	Naïve SS	&	0.1	&	4.01	&	3.99	&	1.00	&	4.0	&	95	&		&		\\	
	&	Bridging MS	&	0.0	&	5.83	&	5.90	&	0.99	&	5.9	&	95	&	0.0	&	94	\\	
	&	Bridging SS	&	0.1	&	3.30	&	3.34	&	0.99	&	3.3	&	94	&		&		\\	\\
2	&	Naïve MS	&	-4.6	&	4.63	&	4.54	&	1.02	&	6.5	&	83	&	14.0	&	3	\\	
	&	Naïve SS	&	9.4	&	3.02	&	2.99	&	1.01	&	9.8	&	14	&		&		\\	
	&	Bridging MS	&	-0.2	&	7.11	&	7.10	&	1.00	&	7.1	&	95	&	0.3	&	94	\\	
	&	Bridging SS	&	0.1	&	4.31	&	4.29	&	1.00	&	4.3	&	94	&		&		\\	\\
3	&	Naïve MS	&	11.6	&	4.93	&	4.80	&	1.03	&	12.6	&	35	&	-2.2	&	92	\\	
	&	Naïve SS	&	9.4	&	3.02	&	2.99	&	1.01	&	9.8	&	14	&		&		\\	
	&	Bridging MS	&	19.7	&	7.10	&	7.06	&	1.01	&	20.9	&	22	&	-19.6	&	2	\\	
	&	Bridging SS	&	0.1	&	4.31	&	4.29	&	1.00	&	4.3	&	94	&		&		\\	\\
4	&	Naïve MS	&	6.8	&	4.30	&	4.23	&	1.02	&	8.0	&	65	&	14.0	&	3	\\	
	&	Naïve SS	&	20.8	&	2.49	&	2.45	&	1.01	&	20.9	&	0	&		&		\\	
	&	Bridging MS	&	16.3	&	6.40	&	6.42	&	1.00	&	17.6	&	29	&	0.3	&	94	\\	
	&	Bridging SS	&	16.7	&	3.48	&	3.52	&	0.99	&	17.0	&	1	&		&		\\	\\
5	&	Naïve MS	&	13.1	&	4.56	&	4.47	&	1.02	&	13.8	&	18	&	7.6	&	47	\\	
	&	Naïve SS	&	20.8	&	2.49	&	2.45	&	1.01	&	20.9	&	0	&		&		\\	
	&	Bridging MS	&	25.3	&	6.45	&	6.44	&	1.00	&	26.1	&	3	&	-8.7	&	57	\\	
	&	Bridging SS	&	16.7	&	3.48	&	3.52	&	0.99	&	17.0	&	1	&		&		\\

\hline
\end{tabular}
\begin{tablenotes}
      \item ASE=average estimated standard error; ESE=empirical standard error; SER=standard error ratio (ASE/ESE); RMSE = root mean squared error; CI = confidence interval; SS=single-span; MS=multi-span;  Monte Carlo standard error for 95\% CI coverage was 0.5\% when coverage was 95\%.
    \end{tablenotes}
  \end{threeparttable}
\end{table}

\begin{table}
\begin{threeparttable}
\caption{Simulation summary results for continuous outcome, $n_1=1000$, $n_2=2000$, $2000$ simulations. Bias, ASE, ESE, SER, RMSE, and 95\% CI coverage calculated for the ATE.}
\label{tab:appn11000cont}
\centering
\setlength{\tabcolsep}{3.5pt} 

\begin{tabular}{c l r r r r r c c c} 
\hline
  Scenario & Estimator & Bias & ASE & ESE & SER & RMSE & \makecell{95\% CI \\  Coverage (\%)} & \makecell{Mean \\ Diagnostic} & \makecell{Diagnostic \\ 95\% CI \\  Includes Zero (\%)}\\ 
  \hline 
1	&	Naïve MS	&	0.0	&	3.29	&	3.37	&	0.98	&	3.4	&	94	&	0.0	&	94	\\	
	&	Naïve SS	&	-0.1	&	2.33	&	2.37	&	0.98	&	2.4	&	94	&		&		\\	
	&	Bridging MS	&	0.0	&	3.29	&	3.35	&	0.98	&	3.4	&	94	&	0.0	&	95	\\	
	&	Bridging SS	&	0.0	&	1.42	&	1.45	&	0.98	&	1.4	&	94	&		&		\\	\\
2	&	Naïve MS	&	17.8	&	4.11	&	4.19	&	0.98	&	18.3	&	1	&	20.3	&	0	\\	
	&	Naïve SS	&	38.1	&	3.30	&	3.34	&	0.99	&	38.3	&	0	&		&		\\	
	&	Bridging MS	&	-0.1	&	4.26	&	4.39	&	0.97	&	4.4	&	94	&	0.1	&	94	\\	
	&	Bridging SS	&	-0.1	&	2.36	&	2.38	&	0.99	&	2.4	&	94	&		&		\\	\\
3	&	Naïve MS	&	37.2	&	4.06	&	4.14	&	0.98	&	37.4	&	0	&	1.0	&	92	\\	
	&	Naïve SS	&	38.1	&	3.30	&	3.34	&	0.99	&	38.3	&	0	&		&		\\	
	&	Bridging MS	&	19.0	&	4.19	&	4.32	&	0.97	&	19.5	&	1	&	-19.0	&	0	\\	
	&	Bridging SS	&	-0.1	&	2.36	&	2.38	&	0.99	&	2.4	&	94	&		&		\\	\\
4	&	Naïve MS	&	47.8	&	4.10	&	4.19	&	0.98	&	48.0	&	0	&	20.3	&	0	\\	
	&	Naïve SS	&	68.1	&	3.29	&	3.34	&	0.99	&	68.2	&	0	&		&		\\	
	&	Bridging MS	&	29.9	&	4.26	&	4.39	&	0.97	&	30.2	&	0	&	0.1	&	94	\\	
	&	Bridging SS	&	29.9	&	2.36	&	2.38	&	0.99	&	30.0	&	0	&		&		\\	\\
5	&	Naïve MS	&	57.8	&	4.10	&	4.19	&	0.98	&	57.9	&	0	&	10.3	&	2	\\	
	&	Naïve SS	&	68.1	&	3.29	&	3.34	&	0.99	&	68.2	&	0	&		&		\\	
	&	Bridging MS	&	39.9	&	4.26	&	4.39	&	0.97	&	40.1	&	0	&	-9.9	&	2	\\	
	&	Bridging SS	&	29.9	&	2.36	&	2.38	&	0.99	&	30.0	&	0	&		&		\\

\hline
\end{tabular}
\begin{tablenotes}
      \item ASE=average estimated standard error; ESE=empirical standard error; SER=standard error ratio (ASE/ESE); RMSE = root mean squared error; CI = confidence interval; SS=single-span; MS=multi-span; Monte Carlo standard error for 95\% CI coverage was 0.5\% when coverage was 95\%.
    \end{tablenotes}
  \end{threeparttable}
\end{table}

\begin{table}
\begin{threeparttable}
\caption{Simulation summary results for binary outcome (formatted as a \%), $n_1=1000$, $n_2=2000$, $2000$ simulations. Bias, ASE, ESE, SER, RMSE, and 95\% CI coverage calculated for the ATE (formatted as a percent).}
\label{tab:appn11000bin}
\centering
\setlength{\tabcolsep}{3.5pt} 

\begin{tabular}{c l r r r r r c c c} 
\hline
  Scenario & Estimator & Bias & ASE & ESE & SER & RMSE & \makecell{95\% CI \\  Coverage (\%)} & \makecell{Mean \\ Diagnostic } & \makecell{Diagnostic \\ 95\% CI \\  Includes Zero (\%)}\\ 
  \hline 

1	&	Naïve MS	&	0.0	&	3.79	&	3.82	&	0.99	&	3.8	&	95	&	-0.1	&	95	\\	
	&	Naïve SS	&	-0.1	&	2.59	&	2.60	&	0.99	&	2.6	&	95	&		&		\\	
	&	Bridging MS	&	0.0	&	3.79	&	3.81	&	0.99	&	3.8	&	95	&	-0.1	&	95	\\	
	&	Bridging SS	&	0.0	&	2.14	&	2.14	&	1.00	&	2.1	&	95	&		&		\\	\\
2	&	Naïve MS	&	-4.5	&	3.10	&	3.12	&	1.00	&	5.5	&	69	&	13.8	&	0	\\	
	&	Naïve SS	&	9.3	&	2.06	&	2.05	&	1.01	&	9.5	&	1	&		&		\\	
	&	Bridging MS	&	-0.2	&	4.62	&	4.68	&	0.99	&	4.7	&	94	&	0.1	&	94	\\	
	&	Bridging SS	&	-0.1	&	2.78	&	2.74	&	1.02	&	2.7	&	96	&		&		\\	\\
3	&	Naïve MS	&	11.7	&	3.27	&	3.30	&	0.99	&	12.1	&	6	&	-2.4	&	84	\\	
	&	Naïve SS	&	9.3	&	2.06	&	2.05	&	1.01	&	9.5	&	1	&		&		\\	
	&	Bridging MS	&	19.6	&	4.61	&	4.67	&	0.99	&	20.1	&	1	&	-19.7	&	0	\\	
	&	Bridging SS	&	-0.1	&	2.78	&	2.74	&	1.02	&	2.7	&	96	&		&		\\	\\
4	&	Naïve MS	&	6.9	&	2.89	&	2.92	&	0.99	&	7.5	&	32	&	13.8	&	0	\\	
	&	Naïve SS	&	20.7	&	1.73	&	1.72	&	1.00	&	20.8	&	0	&		&		\\	
	&	Bridging MS	&	16.4	&	4.19	&	4.23	&	0.99	&	17.0	&	3	&	0.1	&	94	\\	
	&	Bridging SS	&	16.5	&	2.31	&	2.32	&	1.00	&	16.7	&	0	&		&		\\	\\
5	&	Naïve MS	&	13.3	&	3.05	&	3.09	&	0.99	&	13.6	&	1	&	7.4	&	18	\\	
	&	Naïve SS	&	20.7	&	1.73	&	1.72	&	1.00	&	20.8	&	0	&		&		\\	
	&	Bridging MS	&	25.4	&	4.22	&	4.30	&	0.98	&	25.8	&	0	&	-8.9	&	20	\\	
	&	Bridging SS	&	16.5	&	2.31	&	2.32	&	1.00	&	16.7	&	0	&		&		\\	

\hline
\end{tabular}
\begin{tablenotes}
      \item ASE=average estimated standard error; ESE=empirical standard error; SER=standard error ratio (ASE/ESE); RMSE = root mean squared error; CI = confidence interval; SS=single-span; MS=multi-span;  Monte Carlo standard error for 95\% CI coverage was 0.5\% when coverage was 95\%.
    \end{tablenotes}
  \end{threeparttable}
\end{table}

\begin{figure}
\begin{center}
  \includegraphics[width=.85\columnwidth]{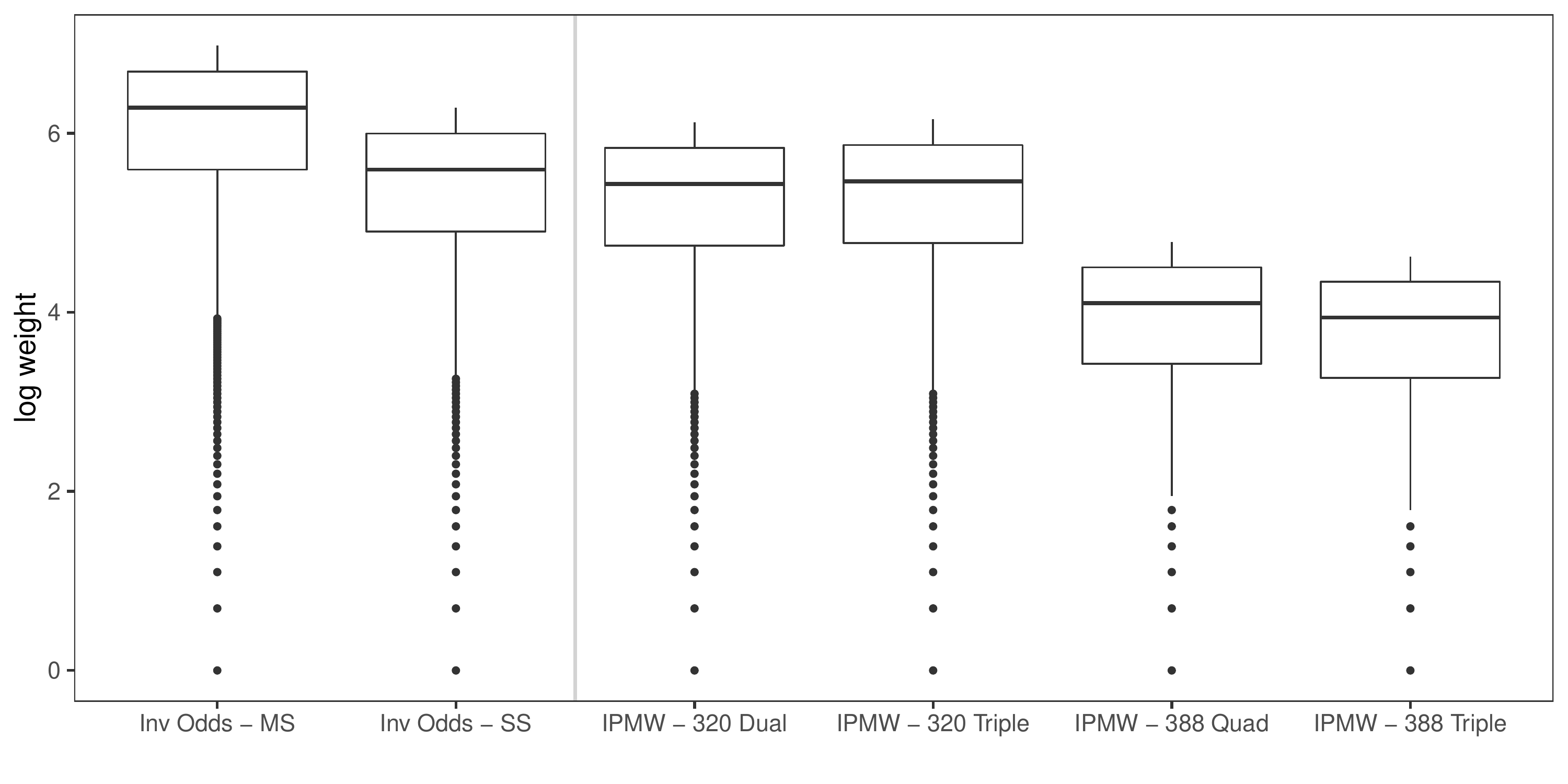}
  \caption{Distribution of the log inverse odds of sampling weights and log inverse probability of missingness weights (IPMW) for the restricted model with baseline CD4 cell counts $< 200$ from Section 4}
  \label{fig:weights}
\end{center}
\end{figure}

	\end{appendices}	
	
	\label{lastpage}
	
\end{document}